\documentclass[10pt,a4paper]{article}

\usepackage{jheppub}

\usepackage[utf8]{inputenc}

\usepackage{natbib}
\usepackage{graphicx}
\usepackage{epsfig}
\usepackage{amsmath}
\input{epsf}
\usepackage{psfrag}

\usepackage[usenames,dvipsnames]{color}

\newcommand{\beq}{\begin{eqnarray}}
\newcommand{\eeq}{\end{eqnarray}}
\newcommand{\Slash}[1]{{\ooalign{\hfil/\hfil\crcr$#1$}}}

\newcommand{\Deq}{\stackrel{\Delta}{\rightarrow}}

\newcommand{\nn}{\nonumber \\}

\usepackage{amsmath,color}
\usepackage{amssymb}
\usepackage{bm}% bold math
\usepackage{graphicx}
\usepackage{multirow}

\newcounter{RSQ}

\begin{document}

\title{Twist analysis of the spin-orbit correlation in QCD
}

\author[a,b]{Yoshitaka Hatta}
\author[b]{Jakob Schoenleber}

\affiliation[a]{Physics Department, Brookhaven National Laboratory, Upton, NY 11973, USA}
\affiliation[b]{RIKEN BNL Research Center, Brookhaven National Laboratory, Upton, NY 11973, USA}
\emailAdd{yhatta@bnl.gov, jschoenle@bnl.gov}

\abstract{We present a QCD analysis of the twist-three parton distribution functions associated with the spin-orbit correlation of quarks and gluons in spin-$\frac{1}{2}$ and spin-0 hadrons. We derive exact non-perturbative identities decomposing the spin-orbit correlations into  the Wandzura-Wilczek part and the genuine twist-three part.  In the spin-$\frac{1}{2}$ case, the result is partially related to the kinematical twist-three part of the $g_T(x)$ distribution familiar in the context of transverse spin physics. We use these identities to obtain a novel longitudinal momentum sum rule which may be regarded as the momentum version of the Jaffe-Manohar spin sum rule. 
 We explore the  physical interpretation of the sum rule and make a connection to the color Lorentz forces and their associated potential energies.}

\maketitle

\section{Introduction} 

Quarks and gluons, fundamental constituents of the nucleon, carry angular momentum in the forms of spin and orbital angular momentum (OAM). These contributions  must sum up to the spin of the nucleon, which is $\frac{1}{2}$, through the process of  angular momentum coupling. This simple argument can be precisely formulated in QCD in terms of spin sum rules    
\beq
\frac{1}{2}&=&\frac{1}{2}\Delta \Sigma + \Delta G + L_q+L_g, \label{jm} \\
&=& \frac{1}{2} \Delta \Sigma + L^{\rm kin}_q+J_g. \label{ji}
\eeq
(\ref{jm}) is referred to as the Jaffe-Manohar (JM) sum rule 
\cite{Jaffe:1989jz}. The quark and gluon helicity contributions $\Delta\Sigma$ and $\Delta G$ are given by the moment 
\begin{align}
\Delta \Sigma =  \sum_q \int_0^1 dx\, (\Delta q(x) + \Delta \bar q(x) ), \quad \Delta G = \int_0^1 dx\, \Delta G(x),
\end{align}
of the polarized quark and antiquark parton distribution functions (PDFs) $\Delta q(x), \Delta \bar q(x)$, and the polarized gluon PDF $\Delta G(x)$. $L_{q,g}$ are the canonical OAMs of quarks and gluons. 
Their corresponding PDFs 
\begin{align}
L_{q} =\sum_q \int_{-1}^1 dx\, L_{q}(x), \qquad L_{g} = \int_{0}^1 dx\, L_{g}(x), 
\end{align}
can be expressed \cite{Lorce:2011kd,Hatta:2011ku} by the moment of the Wigner distributions $f_{q,g}(x,b_{\perp}  k_{\perp})$ \cite{Belitsky:2003nz}
\begin{align}
L_q(x) = \int d^2k_\perp d^2b_\perp \epsilon^{ij}b^i k^j f_q(x,k_\perp,b_\perp),
\label{Lqx def}
\\
xL_g(x) = \int d^2k_\perp d^2b_\perp \epsilon^{ij}b^i k^j f_g(x,k_\perp,b_\perp),\label{Lgx def}
\end{align} 
  which describe the phase space distribution of partons in transverse momentum $k_{\perp}$ space and tranverse position (impact parameter) $b_{\perp}$ space.

%are the corresponding OAM densities, encoding the OAM of quarks and gluons carrying a particular momentum fraction $x$ of the parent hadron. 

An alternative decomposition  (\ref{ji}) is called the Ji sum rule \cite{Ji:1996ek}. While the quark helicity contribution $\Delta\Sigma$ is common,  the  quark kinetic OAM $L_q^{\rm kin}$ is  different from the canonical OAM $L_q$. Moreover, the total gluon angular momentum $J_g$ cannot be decomposed into helicity and OAM parts in a gauge invariant way. The sum rules in (\ref{jm}), (\ref{ji}) shed light on the origin of the nucleon spin, which is one of the main scientific goals of the future Electron-Ion Collider \cite{AbdulKhalek:2021gbh}. Ultimately, one would like to determine the precise numerical values of all the components in (\ref{jm}) and (\ref{ji}) and understand the intricate QCD dynamics behind them.  

Needless to say, the sum rules (\ref{jm}), (\ref{ji}) are relevant only to hadrons with spin-$\frac{1}{2}$. For a spinless hadron, such as the pion, all the terms in  (\ref{jm}), (\ref{ji}) are  zero.  Of course, individual quarks and gluons do carry spin and OAM, but their averages  are zero in a spinless or unpolarized hadron. However,  their {\it correlations} can be  nonzero in general, and may provide novel insights into the structure of hadrons with or without spin.   
By correlations, we mean an analog of spin-orbit coupling ubiquitous in atomic and subatomic physics. For example, it is well known that   the spin-orbit coupling  $\vec{L}\cdot \vec{S}$ of an electron in a hydrogen atom partly contributes to the hydrogen fine structure. Also, in atomic nuclei, the spin-orbit coupling of protons and neutrons orbiting around the nuclear mean field potential is crucial to explain the nuclear magic number \cite{mayer}. 

Similarly, the spin (or helicity) and OAM of quarks and gluons inside a hadron talk to each other. However, the theoretical description of this phenomenon in gauge theories like QCD is challenging. The closest analog of  spin-orbit coupling for quarks and gluons, which we refer to as spin-orbit correlation, was first introduced in \cite{Lorce:2011kd}  and subsequently developed \cite{Kanazawa:2014nha,Mukherjee:2014nya,Lorce:2014mxa,Mukherjee:2015aja,Rajan:2017cpx,Tan:2021osk,Engelhardt:2021kdo}. Similarly to the OAM PDFs $L_{q,g}(x)$, the spin-orbit correlation can be defined as $x$-dependent distributions 
\begin{align}
C_q(x) = \int d^2k_\perp d^2b_\perp \epsilon^{ij}b^i k^j \tilde f_q(x,k_\perp,b_\perp),
\label{Cqx def}
\\
xC_g(x) = \int d^2k_\perp d^2b_\perp \epsilon^{ij}b^i k^j \tilde f_g(x,k_\perp,b_\perp),
\label{Cgx def}
\end{align}
where now $\tilde f_{q,g}(x,k_{\perp}, b_{\perp})$ are the polarized quark and gluon Wigner distributions.
 Intuitively, $C_{q,g}(x)$ can be understood as the product  $L_3S_3$ of the longitudinal components of $\vec{L}$ and $\vec{S}$ summed over all quarks or gluons  with the same value of $x$.  Positive (negative) $C_{q,g}(x)$ indicates that helicity and OAM of partons carrying a momentum fraction $x$ are aligned (anti-aligned).  

It has been recently observed  based on an effective theory of QCD at small-$x$ that the helicity and OAM of small-$x$ quarks and gluons are perfectly anti-aligned even in an unpolarized or spinless hadron \cite{Bhattacharya:2024sno}. Namely, $L_3S_3=-\frac{1}{2}$ for quarks and $L_3S_3=-1$ for gluons, resulting in  the approximate relations
\beq
C_q(x)\approx -\frac{1}{2}q(x), \qquad C_g(x)\approx -G(x), \qquad (x\ll 1) \label{fir}
\eeq
where $q(x),G(x)$ are the unpolarized quark and gluon PDFs. This can be understood as  {\it maximal entanglement} in quantum mechanics between the spin and spatial parts of the wavefunction of each quark and gluon. Additionally, an experimental observable which can test the prediction (\ref{fir}) has been proposed  \cite{Bhattacharya:2024sck}. This opens up   a novel avenue to discuss quantum entanglement effects inside hadrons and nuclei and their manifestation in experimental observables. 

Motivated by these developments, in this paper we perform a QCD analysis of the quark and gluon spin-orbit correlations $C_{q,g}(x)$ for generic values of $x$. To be  precise,  `spin' and `orbit' refer to those in the JM decomposition (\ref{jm}). (In  the Ji decomposition  (\ref{ji}), it is not possible to separately define gluon helicity and gluon OAM.) As we shall discuss extensively,  $C_{q,g}(x)$ do not have a definite twist \cite{Rajan:2017cpx}, similarly to the $g_2(x)$ structure function for the transversely polarized nucleon  \cite{Shuryak:1981pi,Bukhvostov:1983eob,Ratcliffe:1985mp}.  It is well known that such distributions can be uniquely decomposed into the Wandzura-Wilczek (WW) part \cite{Wandzura:1977qf} related to twist-two distributions and the so-called genuine twist-three distributions \cite{Balitsky:1987bk,Geyer:1999uq,Geyer:2000ig}. We shall derive exact formulas \eqref{final1} and \eqref{final2} which completely reveal  the twist structure of $C_{q,g}(x)$, and we check their  consistency with (\ref{fir}) at small-$x$. To a large extent, our discussion is parallel to that in \cite{Hatta:2012cs} where it was shown that the canonical OAM distributions \eqref{Lqx def}, \eqref{Lgx def} can be written entirely in terms of twist-two PDFs and twist-three generalized parton distributions (GPDs). In a sense, $C_{q,g}(x)$ are the parity transforms of $L_{q,g}(x)$, so their operator structures are similar. 
%We show that the analysis can be performed in an analogous fashion to \cite{Hatta:2012cs} to obtain formulas for $C_{q,g}(x)$ in terms of twist-two PDFs and twist-three GPDs, see \eqref{final1} and \eqref{final2}. This is important, because the expression in terms of GTMDs, \eqref{Cqx def} and \eqref{Cgx def} are impractical, since currently we are far from obtaining decent knowledge of GTMDs.  On the other hand the experimental prospects for twist-three GPDs are better, especially in view of the high precision experiments that will be performed for instance the future EIC. 
In the quark sector, our results partly overlap with those in \cite{Rajan:2017cpx}, but we provide more explicit expressions for the genuine twist-three terms. The results in the gluon sector are entirely new.  

Coming back to the sum rule  \eqref{jm},  
%which involves the  first moments $L_{q,g}$ of $L_{q,g}(x)$, 
given the formal analogy between $L_{q,g}$ and $C_{q,g}$, 
%, given by simply replacing $f_{q,g} \leftrightarrow \tilde f_{q,g}$, 
one could ask whether one can derive a similar sum rule involving $C_{q,g}$ instead of $L_{q,g}$, and in fact the answer is positive. As a straightforward corollary to our main results \eqref{final1} and \eqref{final2}, we shall derive a new longitudinal {\it momentum} sum rule 
\eqref{main}. In contrast to \eqref{jm} which only involves the first moments of PDFs, \eqref{main} involves higher moments and explicitly contains genuine twist-three distributions. We shall attempt to provide the physical interpretation of these twist-three terms in terms of potential energies associated with the color Lorentz force.

%The paper is outlined as follows. In Section 2 we give the operators definitions of the spin-orbit correlations $C_{q,g}(x)$. In Section 3, we consider the quark case and derive \eqref{final1} while in Section 4 we deal with the gluon case and derive \eqref{final2}.
%Section 5 is devoted to the new momentum sum rule \eqref{main}. We conclude in Section 6. 

\section{Notation and conventions}
In this section we summarize our notation, conventions and kinematics used throughout. We write a vector in light-cone components $k^{\mu} = (k^+, k_{\perp}, k^-)$, where
\begin{align}
k^{\pm} = \frac{1}{\sqrt{2}} (k^0 \pm k^3), \qquad k_{\perp} = (k^1, k^2).
\end{align}
Latin letters $i,j \in \{ 1,2 \}$ denote transverse indices, while Greek letters denote space-time indices. We contract transverse vectors using the Euclidean signature $k_{\perp} \cdot z_{\perp} = - k^i z_i = k^i z^i$. 

We use the opposite sign convention for $\gamma_5$ compared to \cite{Hatta:2012cs}, namely,  $\gamma_5=i\gamma^0\gamma^1\gamma^2\gamma^3$. The totally anti-symmetric tensor $\epsilon^{\mu\nu\rho\lambda}$ is defined as $\epsilon^{0ij3}=\epsilon^{-+ij}=\epsilon^{ij}$ such that ${\rm Tr}[\gamma^\mu \gamma^\nu\gamma^\rho\gamma^\lambda \gamma_5]=-4i\epsilon^{\mu\nu\rho\lambda}$.  

All operators that appear are sandwiched between off-forward external hadron states. In the case of spin-$\frac{1}{2}$ hadrons, we use light-cone helicity states and average over light-cone helicities
\begin{align}
\frac{1}{2} \sum_{\lambda \in \{+,-\}} \langle p', \lambda | \mathcal O | p , \lambda \rangle.
\end{align}
Spin-averaging will be omitted in the notation, but is always assumed unless otherwise stated.
We define
\begin{align}
P^{\mu} = \frac{p^{\mu} + p'^{\mu}}{2}, \qquad \Delta^{\mu} = p'^{\mu} - p^{\mu}, \qquad t = \Delta^2.
\end{align}
We choose a frame such that $P_{\perp} = 0$ so that
\begin{align}
P^\mu &= \left(P^+,0_\perp, \frac{M^2+\Delta_\perp^2/4}{2P^+(1-\xi^2)} \right),
\\
\Delta^\mu &= \left(-2\xi P^+,\Delta_\perp, \frac{\xi(M^2+\Delta_\perp^2/4)}{P^+(1-\xi^2)}\right),
\end{align}
where $P^+ > 0$, $M$ is the mass of the hadron and $\xi = -\frac{\Delta^+}{2P^+}$ is the skewness. For our purposes we can set $\xi = 0$, hence the $\xi$-dependence in the argument of functions will be omitted throughout this work.
On the other hand, $\Delta_\perp$ is assumed to be small but nonzero. We are chiefly interested in  the first $\frac{\partial}{\partial \Delta_{\perp}}$ derivative of the quantities involved. To this end, we define the symbol $\Deq$ by
\begin{align}
f(\Delta_{\perp}) \Deq \Big ( \frac{\partial f}{\partial \Delta_{\perp}^{\nu}} \Big |_{\Delta_{\perp} = 0} \Big ) \Delta_{\perp}^{\nu},
\label{Deqdef}
\end{align}
i.e. we keep only the linear term in $\Delta_{\perp}$.

The covariant derivative is defined as $D^\mu=\partial^\mu+igA^\mu$. We also use the notations
\begin{align}
\overleftarrow{D}^\mu = \overleftarrow{\partial}^\mu -igA^\mu, \qquad \overleftrightarrow{D}^\mu = \frac{D^\mu -\overleftarrow{D}^\mu}{2}.
\end{align}  
With this convention, the Yang-Mills equation reads $D_\mu F^{\mu\nu}_a=\sum_q g\bar{q}\gamma^\nu  t^aq$. The straight-Wilson line in the fundamental representation  between the space-time positions $x$ and $y$ is denoted by
\begin{align}
W_{x,y} = P \exp\Big ( - ig \int_0^1 ds \, (y-x)^{\mu} A_{\mu}^a((1-s)x +s y) t^a \Big ).
\end{align}
We denote by $\mathcal W_{x,y}$ the straight Wilson in the adjoint representation.

We will often use the light-cone vector in units of position
\begin{align}
n^{\mu} = \frac{1}{\sqrt{2} P^+} (1,0,0,-1)^{\mu},
\end{align}
which projects a vector $k^{\mu}$ onto its plus component divided by $P^+$, i.e. $n \cdot k = k^+/P^+$.

%For $x$-moments of various PDFs $f(x)$, we introduce the notation
%\begin{align}
%    f^{(n)} = \int dx \, x^{n-1} f(x).
%\end{align}
%If integration limits are not shown, it is implied that we are integrating from $-\infty$ to $\infty$. 

Starting from Section \ref{sec: Quark spin-orbit correlation} we will simplify the notation as follows:
\begin{itemize}
    \item We omit the space-time point indices on Wilson lines. They can be easily restored by demanding that the operators are gauge invariant. 
    \item We omit the $\Delta_{\perp}$ dependence in the arguments of functions, which is usually set to zero. Since we are always interested in the linear term in $\Delta_{\perp}$, we can effectively drop terms of order $O(\Delta_{\perp}^2)$. 
\end{itemize}

\section{Spin-orbit correlation and GTMD}

In this section, we introduce the quark and gluon spin-orbit correlations as certain moments of generalized transverse momentum dependent distributions (GTMDs) which are the Fourier transforms of the Wigner distributions 
\beq
\tilde{f}_{q,g}(x,k_\perp,b_\perp)=\int \frac{d^2\Delta_\perp}{(2\pi)^2} e^{-ib_\perp\cdot \Delta_\perp}\tilde{f}_{q,g}(x,k_\perp,\Delta_\perp),
\eeq
Our analysis mostly focuses on the nucleon, or more generally, spin-$\frac{1}{2}$ hadrons, but we also occasionally mention spin-0 hadrons such as the pions or the $\,^4{\rm He}$ nucleus.  

Let us first consider the polarized quark GTMD for the nucleon \cite{Meissner:2009ww}
\beq
&&\tilde{f}_q(x,\xi,k_\perp,\Delta_\perp)= \int \frac{dz^-d^2z_\perp}{2(2\pi)^3} e^{ixP^+z^--ik_\perp\cdot z_\perp}\langle p',s'|\bar{q}(-z/2)W_\pm \gamma^+\gamma_5q(z/2)|p,s\rangle |_{z^+ = 0}\nn 
&& \qquad = \frac{-i}{2M}\bar{u}(p',s')\left[  \frac{\epsilon_{ij}k_\perp^i \Delta_\perp^j}{M^2}G^q_{1,1}+\frac{\sigma^{i+}\gamma_5}{P^+}(k_\perp^i G^q_{1,2}+\Delta_\perp^i G^q_{1,3})+\sigma^{+-}\gamma_5G^q_{1,4}\right]u(p,s).
\label{gtmdqtilde} 
\eeq
Note that the word `polarized' only refers to the operator structure $\bar{q}\gamma^+\gamma_5q$ and does not refer to the spin state of the nucleon. As already mentioned we average over the spin states in the following.
The two possible Wilson lines that connect the quark fields are
\beq
W_\pm\equiv \lim_{w^- \rightarrow \pm \infty} W_{-\frac{z}{2}, (0,- \frac{z_{\perp}}{2}, w^-)}W_{(0,- \frac{z_{\perp}}{2}, w^-),(0,+ \frac{z_{\perp}}{2}, w^-)}W_{(0,+ \frac{z_{\perp}}{2}, w^-),\frac{z}{2}} \Big |_{z^+ = 0}, 
\label{stapleWline}
\eeq
 which connect the two points $\pm z/2$ (with $z^+ = 0$) via light-cone infinity $z^-=\pm \infty$. As we shall see, the direction of the Wilson line $W_+$ or $W_-$ does not matter for the quantities of interest due to $PT$ symmetry \cite{Hatta:2011ku}. Therefore, we suppress the symbol $\pm$ on the left hand side of (\ref{gtmdqtilde}).

In terms of the GTMD, the quark spin-orbit correlation  reads \cite{Lorce:2011kd}
\beq
C_q&=& \left. -i\int_{-1}^1 dx \int dk_\perp \epsilon^{ij}  \frac{\partial}{\partial \Delta^i} k^j \tilde{f}_q(x,k_\perp,\Delta_\perp) \right|_{\Delta_\perp=0} \nn
&=& \int_{-1}^1 dx\int d^2k_\perp  \frac{k_\perp^2}{M^2}G_{1,1}^q(x,k_\perp,0). \label{cqdef}
\eeq
The corresponding PDF is obtained by undoing the $x$-integral  
\beq
C_q(x)=\int d^2k_\perp  \frac{k_\perp^2}{M^2}G_{1,1}^q(x,k_\perp,0).  \qquad (0<x<1) \label{cqx}
\eeq
For antiquarks, $C_{\bar{q}}(x)= -C_q(-x)$ such that
$C_q=\int_0^1dx (C_q(x)-C_{\bar{q}}(x))$. The position space formula (\ref{Cqx def}) is more intuitive as it features  the classical expression of OAM $\epsilon^{ij}b^i k^j$. However, the $\Delta_\perp$-space is more convenient for practical calculations. 
%The physical meaning of this is clearer 
%in terms of the (polarized) Wigner distribution \cite{Belitsky:2003nz} related to the GTMD via Fourier transform  \beq
%\tilde{f}_q(x,k_\perp,b_\perp)=\int \frac{d^2\Delta_\perp}{(2\pi)^2} e^{-ib_\perp\cdot \Delta_\perp}\tilde{f}_q(x,k_\perp,\Delta_\perp),
%\eeq
%where $b_\perp$ is impact parameter. 
%In terms of this, (\ref{cqdef}) can be equivalently written as 
%\beq
%C_q=\int dx \int d^2k_\perp d^2b_\perp \epsilon^{ij}b^i k^j \tilde{f}_q(x,k_\perp,b_\perp).
%\eeq
%This is a more intuitive formula which features the classical expression of OAM $\epsilon^{ij}b^i k^j$. Together with the operator $\tilde{f}_q\sim \bar{q}\gamma^+\gamma_5q$ representing the quark helicity, the physical interpretation of $C_q$ alluded to in the Introduction follows.   

To better understand the operator structure of $C_q(x)$, we convert $k^j$ in (\ref{cqdef}) into a $z_\perp$-derivative acting on the quark bilinear.  
\begin{align} \notag
\int d^2k_\perp k^j \tilde{f}_q &= \frac{i}{2}\int \frac{dz^-}{4\pi}e^{ixP^+z^-}
\\ \notag
&\quad \times \langle p'|\bar{q}(-z/2)\gamma^+\gamma_5\Big [W_{-\frac{z}{2},\frac{z}{2}}D^j_{\rm pure}\Big (\frac{z}{2}\Big)-\overleftarrow{D}^j_{\rm pure}\Big ( -\frac{z}{2} \Big )W_{-\frac{z}{2},\frac{z}{2}} \Big ] q(z/2)|p\rangle \Big|_{z_\perp=0, z^+ = 0}
\\
&\Deq - \frac{i}{2}\epsilon^{ji}\Delta^i C_q(x), \label{ftildem}
\end{align}
where $D^\mu_{\rm pure}$ is the gauge covariant generalization of the partial (canonical) derivative $\partial^\mu$
\begin{align} \notag
D^\mu_{\rm pure}(z) &\equiv D^\mu(z) +i\int dw^- {\cal K}(w^- - z^-)
\\
&\quad \times W_{z,(z^+,z_{\perp},w^-)} gF^{+\mu}(z^+,z_{\perp},w^-)W_{(z^+,z_{\perp},w^-),z}, \label{long} 
\end{align}
with ${\cal K}(w^-)=\pm \theta(\pm w^-)$ ($\theta$ is the Heaviside function) corresponding to the two choices $W_\pm$.  The additional term in (\ref{long}) comes from the derivative of the Wilson line along the light-cone. We recall that $D^\mu_{\rm pure}$ defines the canonical OAM  necessary for the gauge invariant completion of  the Jaffe-Manohar sum rule (\ref{jm}) \cite{Hatta:2011zs,Hatta:2011ku} 
\beq
\langle p',s|\bar{q}\gamma^+i\overleftrightarrow{D}^i_{\rm pure}q|p,s\rangle \Deq is^+ \epsilon^{ij}\Delta_j L_q,
\eeq
where 
\beq
s^\mu=\frac{1}{2}\bar{u}(p,s)\gamma^\mu \gamma_5 u(p,s),
\eeq
is the nucleon spin vector. The same operator naturally appears here since we use the staple-shaped Wilson line.

The gluon spin-orbit correlation can be introduced in the same way. Starting from the polarized gluon GTMD
\beq
x\tilde{f}_g(x,\xi,k_\perp,\Delta_\perp)&=& i\int \frac{d^3z}{(2\pi)^3P^+} e^{ixP^+z^--ik_\perp\cdot z_\perp}\langle p'|\tilde{F}^{+\mu}(-z/2){\cal W}_{\pm} F^{+}_{\ \mu}(z/2)|p\rangle \label{gtmd2}\\ 
&=& \frac{-i}{2M}\bar{u}(p')\left[ \frac{\epsilon_{ij}k^i \Delta^j}{M^2}G^g_{1,1}+\frac{\sigma^{i+}\gamma_5}{P^+}(k^i G^g_{1,2}+\Delta^i G^g_{1,3})+\sigma^{+-}\gamma_5 G^g_{1,4}\right]u(p),  \nonumber
\eeq 
where ${\cal W}_{\pm}$ is the staple-shaped Wilson line, defined in \eqref{stapleWline}, in the adjoint representation, 
we  define the PDF of the gluon spin-orbit correlation as \cite{Kanazawa:2014nha,Bhattacharya:2024sno,Bhattacharya:2024sck}
\beq
xC_g(x)&\equiv &\left.  -i\int d^2k_\perp \epsilon^{ij}   \frac{\partial}{\partial \Delta^i} k^j x\tilde{f}_g(x,k_\perp,\Delta_\perp) \right|_{\Delta_\perp=0} \nn
&=&\int d^2k_\perp  \frac{k_\perp^2}{M^2}G_{1,1}^g(x,k_\perp,0). \label{cgx}
\eeq
Again the directions $\pm$ do not matter. Note that  $C_g(x)$ is odd under $x\to -x$ so the first moment vanishes $\int_{-1}^1 dx C_g(x)=0$ 
in contrast to the quark case. (Incidentally, the integral $\int^1_0 dx C_g(x)$ is divergent.)  Again, eliminating $k_\perp$, we find 
\begin{align} \notag
\int d^2k_\perp k^j x\tilde{f}_g &=  -\frac{1}{2}\int \frac{dz^-}{2\pi P^+} e^{ixP^+z^-} 
\\ \notag
&\quad \times \langle p'|\tilde{F}^{+\mu}\Big ( - \frac{z}{2} \Big )\Big ({\cal W}_{-\frac{z}{2},\frac{z}{2}}\mathcal D^j_{\rm pure} \Big ( \frac{z}{2} \Big )-\overleftarrow{\mathcal D}^j_{\rm pure}\Big (- \frac{z}{2} \Big ) {\cal W}_{-\frac{z}{2},\frac{z}{2}} \Big )F^+_{\ \mu} \Big ( \frac{z}{2} \Big )|p\rangle \Big|_{z_\perp=0,z^+ = 0}
\\
&\Deq - \frac{i}{2}\epsilon^{ji}\Delta^i xC_g(x), \label{gluonC}
\end{align}
where $\mathcal D^j_{\rm pure}$ is $D^j_{\rm pure}$ in the adjoint representation. 

The above discussion for spin-$\frac{1}{2}$ hadrons can be trivially generalized to spin-0 hadrons.  The polarized GTMDs contain just one term  in this case 
\cite{Meissner:2008ay}
\beq
&&\tilde{f}_{q,g}(x,\xi,k_\perp,\Delta_\perp)= -i \frac{\epsilon_{ij}k_\perp^i \Delta_\perp^j}{M^2}G^{q,g}_{1,1}(x,\xi,k_\perp,\Delta_\perp).
\eeq
The other formulas in this section remain the same. 
(\ref{ftildem}) and (\ref{gluonC})  are useful representations of $C_{q,g}(x)$ written solely in terms of collinear (independent of $k_\perp,z_\perp$) operators. These are the starting point of our operator analysis in the next sections.

\section{Quark spin-orbit correlation}
\label{sec: Quark spin-orbit correlation}

%The definition of the spin-orbit correlations  in terms of the Wigner/GTMD distributions is intuitive but somewhat heuristic, as it potentially overlooks the subtleties in properly defining transverse momentum dependent distributions beyond leading order. In order to avoid this issue, it is more advantageous to  work in a purely collinear framework. 
It is known that nonlocal light-ray operator products such as in (\ref{ftildem}) and (\ref{gluonC}) do not carry a definite twist.  This is inconvenient for certain purposes, for example, when considering their scale evolution. In this section, we decompose the quark spin-orbit correlation (\ref{ftildem}) into matrix elements of operators with a definite twist. The general framework for carrying out such a decomposition is well documented in the literature   \cite{Balitsky:1987bk,Geyer:1999uq,Geyer:2000ig}.  
In practice, we closely follow the strategy employed in \cite{Hatta:2012cs} (see also \cite{Ji:2012ba}) for the analysis of the parton OAM PDFs $L_{q,g}(x)$.

\subsection{Twist-3 quark-gluon correlation functions}

As a preliminary, we introduce the `genuine twist-three'  off-forward quark-gluon distributions. First, the  `F-type' correlators are parametrized as\footnote{We use the common notation that for some tensor $T^{\mu ...}$ and some vector $v^{\mu}$, we write $v_{\mu} T^{\mu ...} \equiv T^{v ...}$.}
\beq 
&& \int \frac{d\lambda d\tau}{(2\pi)^2} e^{i\frac{\lambda}{2}(x_1+x_2)+i\tau(x_2-x_1)}\langle p'|\bar{q}(-\lambda n/2)W\Slash n \gamma_5  gF^{n i}(\tau n)Wq(\lambda n/2)|p\rangle \nn
&& = \frac{i}{2} \bar{u}(p')\gamma^i\gamma_5 u(p)\tilde{G}_{Fq}(x_1,x_2)-\frac{\epsilon^{ij}\Delta_j }{2}\bar{u}(p')\Slash n u(p) \tilde{\Lambda}_q(x_1,x_2)
-\frac{i\Delta^i}{2} \bar{u}(p')\Slash n\gamma_5u(p)\tilde{\Phi}_{Fq}(x_1,x_2)+\cdots
\nn
&& \Deq -\epsilon^{ij}\Delta_{ j}\tilde{\Psi}_{Fq}(x_1,x_2),
\label{f1}
\eeq
where $...$ denote further terms that are not relevant for our purposes.
%Since, from this section on, we mostly deal with collinear operators, it is convenient to introduce a null vector $n^\mu=\delta^\mu_-/P^+$ and simplify notations such as  $\gamma^+/P^+=\Slash n$, $F^{+i}/P^+\equiv F^{ni}$.  Also the integration  variables  $\lambda,\tau$ are now   dimensionless. 
The spinor products in the second line have been identified by the following consideration. When $\Delta^+=0$ and hence also $\Delta^-=0$, one has the relations 
\beq 
2P^+\epsilon^{ij}\bar{u}(p')\gamma_j u(p) &=& -i\Delta^i \bar{u}(p')\gamma^+\gamma_5u(p), \label{identi}\\
\epsilon^{ij}\Delta_j \bar{u}(p')\gamma^+u(p) &=& -2iP^+\bar{u}(p')\gamma^i\gamma_5u(p) + 2im \epsilon^{ij}\bar{u}(p')\sigma^+_{\ j} u(p), \\
\bar{u}(p')\gamma^+u(p)
&=&\frac{P^+}{m}\bar{u}(p')u(p)+\frac{i}{2m}\bar{u}(p')\sigma^{+i}\Delta_i u(p), \\
\bar{u}(p')\gamma^i u(p) &=&\bar{u}(p') \frac{i\sigma^{ij}\Delta_j}{2m}u(p),  \label{identities}
\eeq
which follow from the Dirac equation and the Gordon identities. 
There are thus four independent spinor structures with one transverse index 
\beq
i\Delta^i \bar{u}(p')\gamma^+\gamma_5u(p), \quad \epsilon^{ij}\Delta_j \bar{u}(p')\gamma^+u(p), \quad i\bar{u}(p')\gamma^i\gamma_5u(p), \quad \epsilon^{ij}\Delta_j \bar{u}(p')u(p).
\eeq
Since we restrict to linear order in $\Delta_\perp$, the last one is redundant.  
In the last line of (\ref{f1}), we have defined 
\beq
\tilde{\Psi}_{Fq}(x,x')\equiv \lim_{\Delta_{\perp} \rightarrow 0} \left (\frac{1}{2}\tilde{G}_{Fq}(x,x')+\tilde{\Lambda}_q(x,x') \right ) . \label{gf1}
\eeq
The function $\tilde{\Phi}_F$
is related to the quark OAM in a longitudinally polarized nucleon and was the focus of \cite{Hatta:2012cs}. In the present work, this term drops out because we average over nucleon spins, so that $\bar u(p') \gamma^+ \gamma_5 u(p) = 0$, as mentioned before.
Note that, in principle, the distributions depend on $\xi$ and $t$. But one should recall that we have set $\xi=0$ and $\tilde \Psi_F$ does not depend on $t$ by definition. 
For a later use, we equivalently rewrite (\ref{f1}) as 
\beq 
 \int \frac{d\lambda d\tau}{(2\pi)^2} e^{i\frac{\lambda}{2}(x_1+x_2)+i\tau(x_2-x_1)}\langle p'|\bar{q}(-\lambda n/2)W\Slash n   \gamma_5 ig\tilde{F}^{n i}(\tau n)Wq(\lambda n/2)|p\rangle  \Deq i\Delta^i\tilde{\Psi}_{Fq}(x_1,x_2). \label{qgq2}
\eeq
The inverse transform is given by 
\beq
\langle p'|\bar{q}(\zeta n)W\Slash n \gamma_5 ig\tilde{F}^{ni}(\tau n)W q(\lambda n)|p\rangle \Deq i\Delta^i
\int dx_1dx_2 e^{-ix_1\lambda+ix_2\zeta-i(x_2-x_1)\tau}\tilde{\Psi}_{Fq}(x_1,x_2). \label{inverse}
\eeq

Note that $\tilde{G}_F(x_1,x_2)$ is one of the twist-three quark-gluon correlation functions  relevant to transverse single spin asymmetry (SSA) \cite{Efremov:1984ip,Qiu:1991pp,Eguchi:2006qz,Benic:2019zvg}. (The present notation is the same as in \cite{Eguchi:2006qz}.) It is also familiar in the context of higher twist effects in polarized Deep Inelastic scattering \cite{Ratcliffe:1985mp}. This is the only term that survives in the forward limit and for a transversely polarized proton. %$\bar{u}\gamma^i \gamma_5 u= 2s^i$. 
In the present kinematics the same spinor product gives rise to a linear term (recall the averaging over light-cone helicities)
\begin{align}
\bar{u}(p')\gamma^i\gamma_5u(p) = i\epsilon^{ij}\Delta_j  ,  
\label{spinor id}
\end{align}
in momentum transfer.  It is tempting to assume that $\tilde{\Lambda}=0$, in which case no new distribution is introduced. However, we do not see valid reasons to neglect $\tilde{\Lambda}$ in general.

Another correlator without a $\gamma_5$ is 
\beq
&& \int \frac{d\lambda d\tau}{(2\pi)^2} e^{i\frac{\lambda}{2}(x_1+x_2)+i\tau(x_2-x_1)}\langle p'|\bar{q}(-\lambda n/2)W\Slash n gF^{n i}(\tau n) Wq(\lambda n/2)|p\rangle \nn
&& = -\frac{\epsilon^{ij}}{2}\bar{u}(p')\gamma_j \gamma_5u(p) G_{Fq}(x_1,x_2) +    \frac{i\Delta^i }{2}\bar{u}(p')\Slash n u(p) \Lambda_q(x_1,x_2)+\frac{\epsilon^{ij}\Delta_j}{2}\bar{u}(p')\Slash n\gamma_5 u(p) \Phi_{Fq}(x_1,x_2)  +\cdots \nn   
&& \Deq  i\Delta^i\Psi_{Fq}(x_1,x_2),
\label{f2}
\eeq
where we can identify
\beq
\Psi_{Fq}(x,x')\equiv \lim_{\Delta_{\perp} \rightarrow 0} \left ( \frac{1}{2}G_{Fq}(x,x'  )+\Lambda_q(x,x'  ) \right ).
\label{gf2}
\eeq
Again, we will drop the flavor indices, and note that $\Phi_F$ is the same as in \cite{Hatta:2012cs} and $G_F$ is the other twist-three correlator relevant to SSA. Note that the local matrix element 
\beq
\langle p'|\bar{q}\Slash n  gF^{n i} q|p\rangle \Deq i\Delta^i\int dx_1dx_2\Psi_{Fq}(x_1,x_2) , \label{force}
\eeq
looks familiar in the context of the twist-three correction to the $g_2$ structure function for a transversely polarized nucleon \cite{Shuryak:1981pi} 
\beq
\langle p, s_\perp| \bar{q}\Slash n gF^{n i} q|p, s_\perp\rangle = -\epsilon^{ij}s_j \int dx_1dx_2 G_{Fq}(x_1,x_2 ) = 2d^q_2 \epsilon^{ij}s_j. \label{after}
\eeq
However, in general $d^q_2=-\frac{1}{2}\int dx_1dx_2 G_{Fq}$ is different from $-\int dx_1dx_2\Psi_{Fq}$ because of the $\Lambda$-term.

Next, we introduce the `D-type' correlators
\begin{align} \notag
&\int \frac{d\lambda d\tau}{(2\pi)^2} e^{i\frac{\lambda}{2}(x_1+x_2)+i\tau(x_2-x_1)}\langle p'|\bar{q}(-\lambda n/2)\Slash n\gamma_5 W\overleftrightarrow{D}^i(\tau n)Wq(\lambda n/2)|p\rangle
\\
&\qquad 
\Deq -\epsilon^{ij}\Delta_{ j}\tilde{\Psi}_{Dq}(x_1,x_2) ,
\label{d1}
\\ \notag
&\int \frac{d\lambda d\tau}{(2\pi)^2} e^{i\frac{\lambda}{2}(x_1+x_2)+i\tau(x_2-x_1)}\langle p'|\bar{q}(-\lambda n/2)\Slash n W\overleftrightarrow{D}^i(\tau n)Wq(\lambda n/2)|p\rangle 
\\
&\qquad \Deq i\Delta^i\Psi_{Dq}(x_1,x_2).
\label{d2}
\end{align}
From $PT$ symmetry it follows that 
\beq
\tilde{\Psi}_F(x_1,x_2)=-\tilde{\Psi}_F(x_2,x_1), \qquad \Psi_F(x_1,x_2)=\Psi_F(x_2,x_1), \label{psisym}\\
\tilde{\Psi}_D(x_1,x_2)=\tilde{\Psi}_D(x_2,x_1), \qquad \Psi_D(x_1,x_2)=-\Psi_D(x_2,x_1). 
\eeq
Moreover, by commuting $\overleftrightarrow{D}^i$ with $W$'s (see Appendix \ref{aa} for useful formulas), one can obtain the relation between the $F,D$-type correlators   
\beq
&&\tilde{\Psi}_D(x_1,x_2)=P\frac{1}{x_1-x_2}\tilde{\Psi}_F(x_1,x_2)-\delta(x_1-x_2)C_q(x) ,\label{df1}\\
&& \Psi_D(x_1,x_2)=P\frac{1}{x_1-x_2}\Psi_F(x_1,x_2), 
\label{dd}
\eeq
where $P$ denotes the principal value prescription. 
Note that $C_q$, which was defined in (\ref{ftildem}), appears as the coefficient of the delta function in (\ref{df1}). Indeed, a straightforward calculation gives, in the limit $\Delta_\perp \to 0$, 
\beq
&& \int \frac{d\lambda}{4\pi}e^{i\lambda x} \Biggl\{\langle p'| \bar{q}(-\lambda n/2)\Slash n\gamma_5(WD^i-\overleftarrow{D}^iW)q(\lambda n/2)|p\rangle  \nn
&& \quad + \frac{i}{2}\int d\tau \bigl(\epsilon(\tau-\lambda/2)+\epsilon(\tau+\lambda/2)\bigr)\langle p'|\bar{q}(-\lambda n/2)\Slash n\gamma_5WgF^{n i}(\tau n) Wq(\lambda n/2)|p\rangle   \Biggr\}, \nonumber \\
 &&\Deq\epsilon^{ij}\Delta_{ j} C_q(x), \label{delta}
\eeq 
where $\epsilon(x)=x/|x|$ is the sign function.  See Appendix \ref{aa} for useful formulas to derive this. 
Writing $\epsilon(\tau\pm \lambda/2)=2\theta(\tau\pm \lambda/2)-1=-2\theta(-(\tau\pm \lambda/2))+1$ and removing the $\pm 1$ terms which do not contribute  because $\tilde{\Psi}_F(x,x)=0$, we see that (\ref{delta}) agrees with (\ref{ftildem}). 
Integrating (\ref{delta}) over $x$ gives 
\beq
 \langle p'| \bar{q}\Slash n\gamma_5\overleftrightarrow{D}^iq|p\rangle  + \frac{i}{2}\int d\tau \epsilon(\tau)\langle p'|\bar{q}(0)\Slash n \gamma_5WgF^{n i}(\tau n) Wq(0)|p\rangle \Deq \epsilon^{ij}\Delta_{ j}C_q   . \label{second}
\eeq
In the second term, one can freely replace $\epsilon(\tau)\to \pm 2\theta(\pm \tau)$ without changing the value. The result then agrees with  \cite{Rajan:2017cpx}. (\ref{second}) should be compared to another definition of the quark spin-orbit correlation \cite{Lorce:2014mxa}
\beq
 \langle p'| \bar{q}\Slash n\gamma_5\overleftrightarrow{D}^iq|p\rangle \Deq \epsilon^{ij}\Delta_{ j} C_q^{\rm kin} .  \label{kinoam}
\eeq
$C_q$ and $C_q^{\rm kin}$ are the spin-`orbit' correlations associated with the canonical and kinetic OAMs, respectively. The difference as represented by the second term of (\ref{second}) is  an analog of the `potential orbital angular momentum'  $L_q-L_q^{\rm kin}$, see (\ref{jm}) and (\ref{ji})  \cite{Wakamatsu:2010qj,Hatta:2011ku}. It can be attributed to the different shapes of the Wilson line connecting the two points $z/2$ and $-z/2$ in (\ref{gtmdqtilde}), namely, the staple-shaped \cite{Hatta:2011ku} versus straight \cite{Ji:2012sj} Wilson lines. See \cite{Engelhardt:2021kdo} 
for a recent calculation in lattice QCD. Note that $C_q^{\rm kin}$ does not have a gluonic counterpart while $C_q$ does.  

Finally in this subsection, we mention the well-known properties \cite{Jaffe:1983hp, Diehl:1998sm} that for operators on the light-cone, as used in the operator definition of the $\Psi$'s, the time-ordering is immaterial and hence the ordering of the fields can be changed (respecting Fermi-statistics). Moreover, their support is always limited to $-1\leq x_1,x_2 \leq 1$. This applies equally to the pure gluonic correlators defined later.

\subsection{Equation of motion}

We now derive an exact formula which relates  $C_q(x)$ to generalized parton distributions (GPDs). 
We start with an identity that can be derived from the equation of motion (Dirac equation) $(i\Slash D-m_q)q=\bar{q}(i\overleftarrow{\Slash D}+m_q)=0$   \cite{Balitsky:1987bk,Hatta:2012cs}
\begin{align} \notag
&\bar{q}(-z/2)\gamma^i\gamma_5 \left (W_{-z/2,z/2} D^+(z/2) - \overleftarrow{D}^+(-z/2)W_{-z/2,z/2} \right ) q(z/2) \Big |_{z^+, z_{\perp} = 0}
\\ 
&\quad = \bar{q}\gamma^+\gamma^5(WD^i-\overleftarrow{D}^iW)q +i\epsilon^{ij}\bar{q}\gamma_j(WD^+ +\overleftarrow{D}^+W)q \label{braun}  
\\ \notag
&\quad  -i\epsilon^{ij}\bar{q}\gamma^+(WD_j+\overleftarrow{D}_jW)q -2im_q\epsilon^{ij}\bar{q}\sigma^{+}_{\ j}Wq , 
\end{align}
where $m_q$ is the quark mass and $\sigma^{\mu\nu}=\frac{i}{2}[\gamma^\mu,\gamma^\nu]$. On the right hand side we have omitted the space-time arguments and the $|_{z^+, z_{\perp} = 0}$ to simplify the notation, assuming that they should be clear from the context. The right-most and left-most fields are always separated by $z^{\mu}$ and the arguments of Wilson lines and covariant derivatives are determined by gauge invariance. In the end the limit $z^+, z_{\perp} \rightarrow 0$ shall be taken. We will continue using this convenient notation from now on. Note also that, since we always care about the matrix elements $\langle p' | ... | p \rangle$ and we set $z^+, z_{\perp} = 0$, overall translations will produce a phase $e^{iz^- \Delta^+}$, which is unity for $\xi = 0$.

Equation \eqref{braun} can be easily derived by combining formulas like 
\beq
0&=&\gamma^\mu\gamma^\nu (\Slash D+im_q)q\nn
&=&(g^{\mu\nu}\Slash D-D^\mu \gamma^\nu+D^\nu \gamma^\mu+i\epsilon^{\mu\nu\rho\lambda}D_\rho \gamma_\lambda \gamma_5+im_q\gamma^\mu \gamma^\nu)q.
\eeq
Taking the off-forward matrix element $\langle p'|...|p\rangle$ of (\ref{braun}), we find 
\beq
&& 2\frac{\partial}{\partial z^-} \langle p'|\bar{q}(- z^-/2)\gamma^i\gamma_5 Wq(z^-/2)|p\rangle = \langle p'|\bar{q}\gamma^+\gamma^5(W\overleftrightarrow{D}^i+\overleftrightarrow{D}^iW)q|p\rangle  +i\epsilon^{ij}{\frak D}^+\langle p'|\bar{q}\gamma_jWq|p\rangle \nn 
&& \qquad  -i\epsilon^{ij}\langle p'|\bar{q}\gamma^+(W\overleftrightarrow{D}_j-\overleftrightarrow{D}_jW)q|p\rangle -i\epsilon^{ij} \langle p'|\partial_j(\bar{q}\gamma^+Wq)|p'\rangle -2im_q\epsilon^{ij}\langle p'|\bar{q}\sigma^+_{\ j}Wq|p\rangle , \label{iden}
\eeq
where 
\beq
{\frak D}_\mu f(-x,x)=\lim_{a^\mu\to 0} \frac{1}{a^\mu}\left( f(-x+a,x+a)-f(-x,x)\right), \label{trans}
\eeq
denotes the translation derivative. 
Since we assume $\xi=0$, the term involving
${\frak D}^+$ vanishes, since it will be proportional to $\Delta^+=0$.  The left hand side of (\ref{iden}) can be expressed by  
 the polarized quark GPDs
\beq
&&\langle p'|\bar{q}(-\lambda n/2)\gamma^\mu\gamma_5 Wq(\lambda n/2)|p\rangle =\int dx e^{-ix\lambda} \bar{u}(p') \Biggl[ \gamma^\mu\gamma_5\tilde{H}_q + \frac{\gamma_5\Delta^\mu}{2M}\tilde{E}_q +\gamma^\mu_\perp \gamma_5\tilde{G}_q+ \cdots \Biggr]u(p).\nn
\label{quarkgpd}
\eeq
This may not be the standard parameterization (see, e.g., \cite{Kiptily:2002nx}) and requires an explanation. $\tilde{H}_q$ and $\tilde{E}_q$ are the usual twist-two GPDs relevant when $\mu=+$. When $\mu=\perp$, several  twist-three GPDs appear, and  we have collected in $\tilde{G}_q$ all the terms which lead to the spinor product $\bar{u}(p')\gamma^i\gamma_5 u(p) = i\epsilon^{ij}\Delta_j$.
%by using the identities (\ref{identi})-(\ref{identities}). 
In the notation of \cite{Kiptily:2002nx}, we have
$\tilde{G}_q= \tilde{G}_2+2\tilde{G}_4$. The $\cdots$ in (\ref{quarkgpd}) denotes terms that do not contribute to the linear term $\Delta_{\perp}$, i.e. they vanish if $\Deq$ is applied (by the way, this also applies to the $\tilde E_q$ term in (\ref{quarkgpd}), but we have shown it anyways).

Lorentz covariance dictates that  $\int dx \tilde{G}_q(x)=0$.  
The last term in (\ref{iden}) proportional to the quark mass  can be parameterized by the `transversity' GPD \cite{Diehl:2001pm}
\beq
\langle p'|\bar{q}(-\lambda n/2)\sigma^n_{\ j}Wq(\lambda n/2)|p\rangle \Deq -i\frac{\Delta_j}{M}\int dx e^{-ix\lambda}  H_{1q}(x  ), \label{massgpd}
\eeq
where we defined $H_{1q}= 2\tilde H_T^q +  E_T^q$ in the notation of \cite{Diehl:2001pm} (strictly $H_{1q} = H_{1q}(x,\Delta_{\perp} = 0)$ on the right-hand-side of \eqref{massgpd}).

Inserting (\ref{quarkgpd}), (\ref{massgpd}) into (\ref{iden}) and comparing the coefficients of $\epsilon^{ij}\Delta_{j}$,  we find 
\beq
&& x(\Delta q(x)+\tilde{G}_q(x  )) \nn
&&= q(x)-\frac{1}{2}\int dx'\left( \tilde{\Psi}_{Dq}(x,x')+\tilde{\Psi}_{Dq}(x',x)-\Psi_{Dq}(x',x)+\Psi_{Dq}(x,x')\right)  -\frac{m_q}{M}H_{1q}(x )  \nn
&&= q(x)+ C_q(x) -\int dx'P\frac{1}{x-x'}\left(\tilde{\Psi}_{Fq}(x,x')+ \Psi_{Fq}(x,x')\right)  -\frac{m_q}{M}H_{1q}(x  )  ,
\label{find} 
\eeq
 where $\Delta q(x)=\tilde{H}_q(x )$ is the polarized quark PDF and 
 \beq
 q(x)=\int\frac{d\lambda}{4\pi}e^{ix\lambda}\langle p|\bar{q}(0)W\Slash nq(\lambda n)|p\rangle,
 \eeq
 is the unpolarized quark PDF.  Note that the GPD $\tilde{E}_q$ drops out due to the spin average. (Besides, its contribution would be ${\cal O}(\Delta_{\perp}^2)$.)  
 A result equivalent  to (\ref{find}) was previously derived in \cite{Rajan:2017cpx} (see (82) there), although the authors did not rewrite the genuine twist-three part   explicitly in terms of the quark-gluon correlators $ \Psi_F,\tilde{\Psi}_F$. 
 
%\textcolor{blue}{From now on, in order to simplify the notation, we omit the $t = 0$ in the arguments of functions. It should be clear from the context, whether the limit $t = 0$ is taken or not.} 

\subsection{Lorentz invariant relation} 

(\ref{find}) is not yet in the desired form since the twist-three GPD $\tilde{G}_q$ actually contains the WW part related to twist-two distributions.  
We can  eliminate $\tilde{G}_q$ from (\ref{find}) by utilizing a Lorentz invariant relation \cite{Balitsky:1987bk} which we now derive.  For this purpose, we extend  the left hand side of  (\ref{quarkgpd}) covariantly to slightly off the light cone $z^2\neq 0$. On the right hand side, to maintain covariance, we replace  $\gamma_{\perp}^{\mu} \rightarrow \gamma^{\mu} - \frac{\gamma \cdot z }{P \cdot z} P^{\mu} - \frac{\gamma \cdot P}{P \cdot z} z^{\mu}+ P^2 \Slash z z^{\mu}$
\begin{align} \notag
\langle p'|\bar{q}(-z/2)\gamma^\mu\gamma_5 Wq(z/2)|p\rangle &=\int dx e^{-ix P\cdot z} \bar{u}(p') 
\\
&\quad \times \Biggl[ \gamma^\mu(\tilde{H}_q +\tilde{G}_q) -\frac{\gamma\cdot z}{P\cdot z}P^\mu \tilde{G}_q +  P^2 \Slash z z^{\mu} \tilde G_q  + \cdots \Biggr]\gamma_5u(p),
\label{gen}
\end{align}
where the neglected terms $\cdots$ do not contribute in the following. 
Then, taking the $\mu=+$ component and recalling that $P_{\perp} = 0$ in our frame, we immediately find
\begin{align}
\epsilon^{ij} \Delta_j \tilde{G}_q  = \left.\int \frac{d\lambda}{2\pi} (i \lambda) e^{ix\lambda } \frac{\partial}{\partial z_i}  \langle p'|\bar{q}(-z/2)\Slash n\gamma_5 Wq(z/2)|p\rangle \right|_{z^+ = 0, z_{\perp} = 0}.
\end{align}
The $z_i$-derivative can be evaluated  using the formula (\ref{der1}). This leads to 
\begin{align}
\tilde G_q(x) &= -\int dx'\, \frac{d}{dx} \tilde \Psi_{Dq}(x,x') +  \int dx' \, P \frac{1}{x-x'} \Big ( \frac{\partial}{\partial x} - \frac{\partial}{\partial x'} \Big ) \tilde \Psi_{Fq}(x,x') \notag
\\
&= \frac{d}{dx} C_q(x) + 2 \int dx' \, P \frac{1}{(x-x')^2} \tilde \Psi_{Fq}(x,x'), \label{tildeGq expr}
\end{align}
where we  used \eqref{dd} in the second equality. See also (\ref{partial}). Inserting \eqref{tildeGq expr} into the equations of motion relation  \eqref{find}, we obtain the differential equation 
\beq
x \frac{d}{dx} C_q(x) - C_q(x) &=& - x \Delta q(x) + q(x) - \int dx' \, P \frac{1}{x-x'}  \Psi_{Fq}(x,x') \nn 
&&- \int dx' \, P \frac{3x-x'}{x-x'} \tilde \Psi_{Fq}(x,x')  -\frac{m_q}{M}H_{1q}(x  )  .
\label{eq: deq}
\eeq
%where
%\begin{align}
%\Omega_q(x) \equiv - x \Delta q(x) + q(x) + \int dx' \, P \frac{1}{x-x'}  \Psi_{Fq}(x,x') + \int dx' \, P \frac{3x-x'}{x-x'} \tilde \Psi_{Fq}(x,x').
%\end{align}
%Assuming that the that $C_q$ vanishes at $\pm 1$
%eq. \eqref{eq: deq} has the solution 
%\begin{align}
%C_q(x) = - x \int_x^{\epsilon(x)} \frac{dx'}{x'^2} %\Omega_q(x'),
%\end{align}
%where $\epsilon(x)$ is the sign function.
With the boundary condition $C_q(\pm 1)=0$, this can be solved as
\beq
C_q(x) &=&  x\int_x^{\epsilon(x)} \frac{dx'}{x'} \Delta q(x')  -x\int_x^{\epsilon(x)}\frac{dx'}{x'^2}q(x') \nn &&
+x\int_x^{\epsilon(x)}dx_1\int_{-1}^1 dx_2\frac{\tilde{\Psi}_{Fq}(x_1,x_2)}{x_1-x_2}P\frac{3x_1-x_2}{x_1^2(x_1-x_2)} \nn
&& +x\int^{\epsilon(x)}_xdx_1\int_{-1}^1 dx_2\Psi_{Fq}(x_1,x_2)P\frac{1}{x_1^2(x_1-x_2)} 
\nn && 
+\frac{xm_q}{M}\int_x^{\epsilon(x)} \frac{dx'}{x'^2}H_{1q}(x'  ). \label{final1}
\eeq
In Appendix \ref{ab}, we present an alternative derivation of this result which is more in line with the approach taken in  \cite{Hatta:2012cs}. The above derivation is faster, but this alternative method gives another representation of $\tilde{G}_q$.

(\ref{final1}) achieves the original goal of decomposing $C_q(x)$ into distributions with a definite twist. It is therefore one of our main results. It consists of the WW part related to the twist-two PDFs, the genuine twist-three part and a quark mass term. For antiquarks, $C_{\bar{q}}(x)=-C_q(-x)$. 

A somewhat surprising feature of (\ref{final1}) is that the polarized quark PDF $\Delta q(x)$ survives in this relation even though we have systematically averaged over  nucleon spins at every  step of the derivation. It can be traced back to the $\tilde{H}_q$ term in (\ref{quarkgpd}) which survives when $\mu=\perp$ because of \eqref{spinor id}. Similarly, by definition (see (\ref{gf1}), (\ref{gf2})), $\Psi_F$ and $\tilde{\Psi}_F$ also contain the distributions $G_F$ and $\tilde{G}_F$ unique to spin-$\frac{1}{2}$ hadrons. These spin-dependent distributions actually combine to form a known function, namely, the so-called kinematical twist-three distribution $\tilde{g}(x)$ related to the $g_T(x)$ distribution for a transversely polarized nucleon 
\beq
\int \frac{d\lambda}{4\pi}e^{ix\lambda}\langle p, s_\perp|\bar{q}(0)\gamma^\mu_\perp \gamma_5W q(\lambda n)|p, s_\perp\rangle = s^\mu_\perp g^q_T(x).
\eeq
It is well known that $g_T$ admits the following decomposition 
\cite{Ratcliffe:1985mp,Belitsky:1997ay}
\beq
g^q_T(x)=-\frac{1}{2x}\tilde{g}^q(x) -\frac{1}{2x}\int dx'\frac{\tilde{G}_{Fq}(x,x')+G_{Fq}(x,x')}{x-x'}+\frac{m_q}{M}\frac{h^q_1(x)}{x}, \label{gt}
\eeq
into the kinematical twist-three part $\tilde{g}$ and the  genuine twist-three part  $G_F,\tilde{G}_F$ plus a mass term proportional to  the transversity PDF $h_1$. $\tilde{g}$ can be further decomposed as  \cite{Belitsky:1997ay,Eguchi:2006qz}
\beq 
-\frac{\tilde{g^q}(x)}{2x} &=& \int_x^{\epsilon(x)} \frac{dx'}{x'} \Delta q(x')  
+\frac{1}{2}\int_x^{\epsilon(x)}dx_1\int_{-1}^1 dx_2\frac{\tilde{G}_{Fq}(x_1,x_2)}{x_1-x_2}P\frac{3x_1-x_2}{x_1^2(x_1-x_2)} \nn
&& +\frac{1}{2}\int^{\epsilon(x)}_xdx_1\int_{-1}^1 dx_2G_{Fq}(x_1,x_2)P\frac{1}{x_1^2(x_1-x_2)} 
\nn
&& -\frac{m_q}{M}\int_x^{\epsilon(x)}dx' \frac{h_1^q(x')}{x'^2}.
\eeq
We recognize the exact same linear combination of $\Delta q, G_F,\tilde{G}_F$ in (\ref{final1}). Therefore, (\ref{final1})  can be rewritten   as
\beq
C_q(x)&=& -\frac{\tilde{g}^q(x)}{2}   -x\int_x^{\epsilon(x)}\frac{dx'}{x'^2}q(x')  +x\int_x^{\epsilon(x)}dx_1\int_{-1}^1 dx_2\frac{\tilde{\Lambda}_{q}(x_1,x_2 )}{x_1-x_2}P\frac{3x_1-x_2}{x_1^2(x_1-x_2)}
\nn
&& +x\int^{\epsilon(x)}_xdx_1\int_{-1}^1 dx_2\Lambda_{q}(x_1,x_2)P\frac{1}{x_1^2(x_1-x_2)} 
\nn
&&+ \frac{xm_q}{M}\int_x^{\epsilon(x)}\frac{dx'}{x'^2}\left(H_{1q}(x'  )+h_1^q(x')\right), \label{blu} 
\eeq
where $\Lambda_q$ and $\tilde{\Lambda}_q$, introduced in (\ref{f1}) and (\ref{f2}), are intrinsically off-forward distributions.  We thus see that $C_q(x)$ is partly related to $\tilde{g}(x)$. It is however not the dominant part. The  right hand side of (\ref{blu}) will be   dominated by the unpolarized quark distribution $q(x)$, especially at small-$x$ \cite{Bhattacharya:2024sno,Bhattacharya:2024sck}. 
Besides, $\tilde{g}=0$ (and also $h_1=0$)  for spin-0 hadrons. 

Returning to (\ref{final1}), its first moment reads 
\beq
C_q = \int_{-1}^1dx C_q(x)
%&=&  
%\frac{1}{2}\int_{-1}^1 dx x\Delta q(x) -\frac{1}{2}\int_{-1}^1 dx  q(x) \nn
%&& -\frac{1}{2} \int dx dx'  \left(\frac{3x-x'}{(x-x')^2}\Psi_{Fq}(x,x') + P\frac{1}{x-x'}\tilde{\Psi}_{Fq}(x,x')\right) \nn
&=&\frac{1}{2}\int_0^1 dx x(\Delta q(x) - \Delta\bar{q}(x)) -\frac{N_q}{2}
\nn
&& +\int^1_{-1}\frac{ dxdx'}{x-x'}\tilde{\Psi}_{Fq}(x,x')  +\frac{m_q}{2M}H_{1q}^{(1)},  \label{firstm}
\eeq
where  $N_q=\int_0^1 dx (q(x)-\bar{q}(x))$ is the number of valence quarks with flavor $q$ and 
\beq
 H_{1q}^{(n)}\equiv\int_{-1}^1 dx x^{n-1}H_{1q}(x, t= 0).
 \eeq 
Using (\ref{dd}) one can also write (\ref{kinoam}) as 
\beq
C^{\rm kin}_q\equiv -\int dx dx' \tilde{\Psi}_{Dq}(x,x') = \frac{1}{2}\int_0^1 dx x(\Delta q(x) - \Delta\bar{q}(x)) -\frac{N_q}{2}  +\frac{m_q}{2M}H_{1q}^{(1)}, \label{dlo}
\eeq
 in agreement  with \cite{Lorce:2014mxa}.  We are also interested in the second moment 
\beq 
\sum_q C^{(2)}_{q} &\equiv& \sum_q\int_{-1}^1 dx xC_q(x) \nn
 %&=& \frac{1}{3} \int_{-1}^1 dx x^2  \Delta q(x) -\frac{1}{3} \int_{-1}^1 dx xq(x) \nn 
 %&& -\frac{1}{3}\int dxdx'x\left(\frac{3x-x'}{(x-x')^2}\Psi_F(x,x') + P\frac{1}{x-x'}\tilde{\Psi}_F(x,x')\right)  \nn
&=&\frac{1}{3} \int_0^1 dx x^2 \Delta \Sigma(x) -\frac{ 1}{3}\sum_q(A_{q} +A_{\bar{q}})\label{cq1} \\
&&+\int_{-1}^1 dx dx' \sum_q\left[\frac{x}{x-x'}\tilde{\Psi}_{Fq}(x,x') +\frac{1}{6}\Psi_{Fq}(x,x')\right]  + \sum_q\frac{m_q}{3M}H_{1q}^{(2)}, \nonumber 
\eeq
where 
\beq
\Delta\Sigma(x) \equiv \sum_q (\Delta q(x)+\Delta \bar{q}(x)),
\eeq
and 
\beq
A_q=\int_0^1 dx xq(x), \qquad A_{\bar{q}}= \int_0^1 dx x \bar{q}(x), 
\eeq
are the fractions of the nucleon momentum carried by quarks and antiquarks with flavor $q$, respectively.  
Finally, from (\ref{find}), (\ref{firstm}) and (\ref{cq1}), it follows that 
 \beq
 \int_{-1}^1 dx x\tilde{G}_q(x) &=& -C^{\rm kin}_q,
\label{g1} \\
 \int_{-1}^1dx x^2\tilde{G}_q(x)&=& -\frac{2}{3} \int_0^1 dx x^2 (\Delta q(x)+\Delta \bar{q}(x)) +\frac{2}{3}(A_{q} +A_{\bar{q}}) \nn 
&&  - \frac{1}{3}\int_{-1}^1 dxdx'\Psi_{Fq}(x,x')  -\frac{2m_q}{3M}H_{1q}^{(2)}, \label{g2}
\eeq
and $\int_{-1}^1 dx \tilde{G}_q=0$ as mentioned before. 
(\ref{g1}) agrees with \cite{Kiptily:2002nx,Lorce:2014mxa}.  (\ref{g2}) is consistent with \cite{Rajan:2017cpx} but disagrees with \cite{Kiptily:2002nx} where essentially the authors assumed $\int dx_1dx_2\Psi_F=0$, see a comment after (\ref{after}).

\section{Gluon spin-orbit correlation}

\subsection{Twist-3 three-gluon correlation functions}

We now repeat essentially the same procedure for the gluon spin-orbit correlation $C_g(x)$. 
We first introduce the `F-type' genuine twist-three, three-gluon off-forward correlator   
\beq
 && \int \frac{d\lambda d\tau} {(2\pi)^2}e^{i\frac{\lambda}{2}(x_1+x_2)+i\tau(x_2-x_1)} \langle p'|F^{ni}(-\lambda n/2){\cal W}gF^{nj}(\tau n){\cal W}F^{nk}(\lambda n/2)|p\rangle  \nn 
 &&  = -\frac{1}{2}\left(\delta^{ik}\epsilon^{jl} F(x_1,x_2)-\delta^{ij}\epsilon^{kl} F(x_2,x_2-x_1)-\delta^{jk}\epsilon^{il}F(x_1,x_1-x_2) \right) \bar{u}(p')\gamma^l\gamma_5 u(p)\nn
 && \quad +\frac{i}{2} \left(\delta^{ik}\Delta^j\Lambda_G(x_1,x_2) -\delta^{ij}\Delta^k\Lambda_G(x_2,x_2-x_1)-\delta^{jk}\Delta^i \Lambda_G(x_1,x_1-x_2)\right)\bar{u}(p')\Slash nu(p) \nn 
 && \quad -\frac{1}{2}\left( \delta^{ik}\epsilon^{jl}M(x_1,x_2) + \delta^{ij}\epsilon^{kl}M(x_2,x_2-x_1)-\delta^{ik}\epsilon^{jk}M(x_1,x_1-x_2)\right)\bar{u}(p')\Slash n \gamma_5u(p)\Delta_l+\cdots\nn
&&\Deq  i\left(\delta^{ik}\Delta^j N(x_1,x_2)-\delta^{ij}\Delta^k N(x_2,x_2-x_1)-\delta^{jk}\Delta^i N(x_1,x_1-x_2) \right). 
 \label{yoshida}
\eeq 
The required set of spinor products is the same as in the quark case, whereas the Lorentz index structure has been deduced from related studies in the literature \cite{Ji:1992eu,Beppu:2010qn,Hatta:2012jm}.  
 The field strength tensor in the middle should be understood as a matrix in the adjoint representation  $F^{+j}_{ab}=F^{+j}_c(T^c)_{ab}=-if_{abc}F^{+j}_c$. The rest of the color indices are understood to be contracted in the adjoint representation.  
In the last line, we defined  
\beq
N(x_1,x_2)\equiv \lim_{\Delta_{\perp} \rightarrow 0} \left ( -\frac{1}{2}F(x_1,x_2)+ \Lambda_G(x_1,x_2) \right ). \label{lambdag}
\eeq
In the forward limit only the $F(x_1,x_2)$ terms survive for a transversely polarized nucleon \cite{Ji:1992eu,Beppu:2010qn,Hatta:2012jm}. (The present notation follows \cite{Hatta:2012jm}.)   $M(x_1,x_2)$ is relevant to the gluon OAM in a longitudinally polarized nucleon and was the focus of  \cite{Hatta:2012cs}, but it is irrelevant in the present context. 
$\Lambda_G$ is a new distribution that arises when $\Delta_\perp \neq 0$. It is the gluonic analog of  $\tilde{\Lambda}_q$ and $\Lambda_q$ in (\ref{f1}), (\ref{f2}).
From $PT$ and permutation symmetries \cite{Beppu:2010qn},   \beq
F(x_1,x_2)=F(x_2,x_1)=-F(-x_1,-x_2),
\eeq
and the same relations hold for $N(x_1,x_2)$ and $\Lambda_G(x_1,x_2)$.

Contracting $ik$ indices antisymmetrically, we get   
\beq
&& \epsilon^{ik}\int \frac{d\lambda d\tau} {(2\pi)^2}e^{i\frac{\lambda}{2}(x_1+x_2)+i\tau(x_2-x_1)} \langle p'|F^{ni}(-\lambda n/2){\cal W}gF^{nj}(\tau n){\cal W}F^{nk}(\lambda n/2)|p\rangle \nn 
&& =- \int \frac{d\lambda d\tau} {(2\pi)^2}e^{i\frac{\lambda}{2}(x_1+x_2)+i\tau(x_2-x_1)} \langle p'|\tilde{F}^{n\rho}(-\lambda n/2){\cal W}gF^{nj}(\tau n){\cal W}F^{n}_{\ \ \rho}(\lambda n/2)|p\rangle \nn
&& \Deq -i\left( N(x_1,x_1-x_2)-N(x_2,x_2-x_1)  \right)\epsilon^{jl}\Delta_l \nn
&& \equiv -i \tilde{N}_F(x_1,x_2)\epsilon^{jl}\Delta_l, 
\label{nftil}
\eeq
or equivalently, 
\beq
&& i\int \frac{d\lambda d\tau} {(2\pi)^2}e^{i\frac{\lambda}{2}(x_1+x_2)+i\tau(x_2-x_1)} \langle p'|\tilde{F}^{n\rho}(-\lambda n/2){\cal W}\, ig\tilde{F}^{nj}(\tau n){\cal W}F^{n}_{\ \rho}(\lambda n/2)|p\rangle \nn 
&& \Deq i \tilde{N}_F(x_1,x_2)\Delta^j. \label{ggg2}
\label{nftil}
\eeq
$\tilde{N}_F$ defined in the last line has symmetry properties 
\beq
\tilde{N}_F(x_1,x_2)=-\tilde{N}_F(x_2,x_1)  =-\tilde{N}_F(-x_1,-x_2).
\eeq
On the other hand, contracting the $ik$ indices symmetrically, 
\beq
&& \delta^{ik}\int \frac{d\lambda d\tau} {(2\pi)^2}e^{i\frac{\lambda}{2}(x_1+x_2)+i\tau(x_2-x_1)} \langle p'|F^{ni}(-\lambda n/2){\cal W}gF^{nj}(\tau n){\cal W}F^{nk}(\lambda n/2)|p\rangle \nn 
&&\Deq i(2N(x_1,x_2)-N(x_2,x_2-x_1)-N(x_1,x_1-x_2))\Delta^j \nn
&&\equiv iN_F(x_1,x_2)\Delta^j, 
\eeq
 with 
\beq
 N_F(x_1,x_2)&=&N_F(x_2,x_1)=-N_F(-x_1,-x_2).
 \eeq 
$N_F$ and $\tilde{N}_F$ may be regarded as the gluonic counterparts of $\Psi_F$ and $\tilde{\Psi}_F$. However, unlike the latter, $N_F$ and $\tilde{N}_F$ are not independent.  From their definitions in terms of $N(x_1,x_2)$, it follows that  
 \beq
 \tilde{N}_F(x_1,x_2)&=& N_F(x_1,x_1-x_2)-N_F(x_2,x_2-x_1).
\eeq
Essentially the same observation was made in the forward case \cite{Braun:2009mi}.

Next introduce the `D-type' correlator
\beq
&&\int \frac{d\lambda d\tau} {(2\pi)^2}e^{i\frac{\lambda}{2}(x_1+x_2)+i\tau(x_2-x_1)} \langle p'|\tilde{F}^{n\rho}(-\lambda n/2){\cal W}\overleftrightarrow{D}^j(\tau n){\cal W}F^{n}_{\ \ \rho}(\lambda n/2)|p\rangle \nn
&& \Deq i\tilde{N}_D(x_1,x_2)\epsilon^{jl}\Delta_l,
\eeq
which is symmetric $\tilde{N}_D(x_1,x_2)=\tilde{N}_D(x_2,x_1)=\tilde{N}_D(-x_1,-x_2)$.  
A straightforward calculation (using the results from Appendix \ref{aa}) analogous to the quark case reveals that is related to the $F$-type correlator by
\beq
\tilde{N}_D(x_1,x_2)=\frac{\tilde{N}_F(x_1,x_2)}{x_1-x_2} -\frac{1}{2}\delta(x_1-x_2)x_1C_g(x_1). \label{cgdef}
\eeq
Again, note that the spin-orbit correlation $C_g$ appears as the coefficient of the delta function.

\subsection{Equation of motion} 

The twist structure of $C_g(x)$ can be completely determined by the equation of motion (Yang-Mills equation) and the Lorentz invariance relation. We first consider the equation of motion and define 
\beq
J&\equiv & \langle p'|\frac{\partial}{\partial z^-}\left(\tilde{F}^{+\rho}(0){\cal W}F^i_{\ \rho}(z^-) +\tilde{F}^{i\rho}(-z^-){\cal W}F^+_{\ \ \rho}(0)\right)|p\rangle.
\eeq
On one hand, this can be expressed in terms of the polarized gluon GPDs generalized to twist-three 
\beq
&&\langle p'|\tilde{F}^{\alpha\rho}(-z^-/2){\cal W}F^{\beta}_{\ \rho}(z^-/2)+\tilde{F}^{\beta\rho}(-z^-/2){\cal W}F^{\alpha}_{\ \rho}(z^-/2)|p\rangle \nn 
&&=-i \int dx e^{-ixP^+z^-} \bar{u}(p')\left[\tilde{H}_g P^{(\alpha}\gamma^{\beta)}\gamma_5 +\frac{P^{(\alpha}\Delta^{\beta)}}{2M}\gamma_5\tilde{E}_g + x\tilde{G}_g P^{(\alpha}\gamma^{\beta)}_\perp\gamma_5 +\cdots \right]u(p), \label{gluegpd}
\eeq
where $A^{(\alpha}B^{\beta)}=\frac{A^\alpha B^\beta + A^\beta B^\alpha}{2}$. 
$\tilde{H}_g,\tilde{E}_g$ are the standard twist-two helicity GPDs normalized as $\tilde{H}_g(x)=x\Delta G(x)$ in the forward limit. The twist-three $x\tilde{G}_g$ is defined in the same vein as $\tilde{G}_q$ in (\ref{quarkgpd}) and satisfies $\int dx x\tilde{G}_g(x)=0$.  
We find 
\beq 
J\Deq -i \epsilon^{ij} \Delta_j\frac{(P^+)^2}{2}\int dx e^{-ixP^+z^-} x^2(\Delta G(x)+\tilde{G}_g(x))  . \label{j1}
\eeq
On the other hand, by explicitly carrying out the derivative, we obtain
\beq
J&=& \tilde{F}^{+\rho}(0){\cal W}D^+F^i_{\ \rho}(z^-) -\tilde{F}^{i\rho}(-z^-)\overleftarrow{D}^+{\cal W}F^+_{\ \rho}(0) \nn
&= & \tilde{F}^{+\rho}(0){\cal W}(D^iF^+_{\ \rho}(z^-)-D_\rho F^{+i}(z^-)) +(-\tilde{F}^+_{\ \rho}(-z^-)\overleftarrow{D}^i +\tilde{F}^{+i}(-z^-)\overleftarrow{D}_\rho){\cal W}F^{+\rho}(0) \nn
&& +\epsilon^{ij}D_\alpha  F^{\alpha +}(-z^-){\cal W}F^{+}_{\ j}(0) 
\nn
&=& \tilde{F}^{+\rho}(0)({\cal W}D^i -\overleftarrow{D}^i{\cal W})F^+_{\ \rho}(z^-) +  {\mathfrak D}_\rho (-\tilde{F}^{+\rho}{\cal W}F^{+i}+\tilde{F}^{+i}{\cal W}F^{+\rho}) \nn
&& +i\int_0^{z^-}dw^- \tilde{F}^{+\rho}(0){\cal W}gF_\rho^{\ +}(w^-){\cal W}F^{+i}(z^-) -i\int_0^{z^-}dw^- \tilde{F}^{+i}(0){\cal W}gF_\rho^{\ +}(w^-){\cal W}F^{+\rho}(z^-)  \nn
&&  -\tilde{F}^{+i}(0){\cal W}D_\rho F^{+\rho}(z^-) +\epsilon^{ij} D_\alpha F^{\alpha +}(-z^-){\cal W}F^{+}_{\ j}(0) \nn
&=&\tilde{F}^{+\rho}(0)({\cal W}\overleftrightarrow{D}^i +\overleftrightarrow{D}^i{\cal W})F^+_{\ \rho}(z^-) +  {\mathfrak D}_\rho (-\tilde{F}^{+\rho}{\cal W}F^{+i}+\tilde{F}^{+i}{\cal W}F^{+\rho}) \nn
&& -i\int_0^{z^-}dw^- \tilde{F}^{+\rho}(0){\cal W}gF^+_{\ \  \rho}(w^-){\cal W}F^{+i}(z^-) +i\int_0^{z^-}dw^- \tilde{F}^{+i}(0){\cal W}gF^+_{\ \ \rho}(w^-){\cal W}F^{+\rho}(z^-)  \nn
&&  +\sum_q \epsilon^{ij}\Bigl(\bar{q}(z^-)\gamma^+WgF^+_{\ \ j}(0)Wq(z^-) +\bar{q}(-z^-)\gamma^+ WgF^{+}_{\ \ j}(0)W q(-z^-) \Bigr), \label{j2}
\eeq
where for simplicity we omitted the brackets $\langle p'|...|p\rangle$. It is understood that inside this matrix element one can freely translate the $z^-$ coordinate because $\xi=0$. In deriving (\ref{j2}), we used the Yang-Mills equation $D_\mu F^{\mu\nu}_a=\sum_q g\bar{q}\gamma^\nu t^a q$, the Bianchi identity $D_\mu \tilde{F}^{\mu\nu}=0$  and its variant 
\beq
D_{\mu}\tilde{F}_{\nu\lambda} =D_\nu \tilde{F}_{\mu\lambda}+D_\lambda \tilde{F}_{\nu\mu}+ \epsilon_{\mu\nu\lambda\beta}D_\alpha F^{\alpha\beta}.
\eeq
We also used (\ref{der2}) to obtain the translation derivative. This term connects to the unpolarized gluon PDF 
\beq
\langle p'|  {\mathfrak D}_j(-\tilde{F}^{+j}{\cal W}F^{+i}+\tilde{F}^{+i} {\cal W}F^{+j})|p\rangle &=& i\epsilon^{ij}\Delta_j \langle p'|F^{+\mu}{\cal W}F^+_{\ \mu}|p\rangle \nn
&\Deq& -i\epsilon^{ij}\Delta_j (P^+)^2 \int dx e^{-ixP^+z^-}  xG(x), \label{2di}
\eeq
where we used the Schouten identity $\epsilon^{ik}a^j-\epsilon^{jk}a^i= \epsilon^{ij}a^k$, which holds for any two-dimensional vector $a_\perp$. Equating (\ref{j1}) and (\ref{j2}), we obtain 
\beq
\frac{1}{2}x^2(\Delta G+\tilde{G}_g)&=& xC_g(x)+xG(x) -2\int dx' \frac{\tilde{N}_F(x,x')}{x-x'} \nn && -2\int dx' P\frac{N_F(x,x')}{x-x'} -2\sum_q\int dX \Psi_F\left(X+\frac{x}{2},X-\frac{x}{2}\right) .\label{first}
\eeq

 \subsection{Lorentz invariant relation  }

We now eliminate $\tilde{G}_g$ from (\ref{first}) using a Lorentz invariant relation. As in the quark case, we generalize (\ref{gluegpd}) to  off the light-cone in the following way
\beq
&&  \langle p'|\tilde{F}^{\alpha\rho}(-z/2){\cal W}F^\beta_{\ \rho}(z/2)+\tilde{F}^{\beta\rho}{\cal W}F^{\alpha}_{\ \rho}|p\rangle  \label{glu3} \\
&& = -\frac{i}{2}\int dx e^{-ix  P\cdot z}\bar{u}(p')\left(( P^\alpha \gamma^\beta+ P^\beta  \gamma^\alpha)(\tilde{H}_g+x\tilde{G}_g)-2\frac{\gamma\cdot z}{P\cdot z} P^\alpha P^\beta x\tilde{G}_g + \cdots\right)\gamma_5u(p). \nonumber
\eeq
This immediately gives 
\begin{align}
 -2\left. \int \frac{dz^-}{2\pi} e^{ixz^- P^+} z^-  \frac{\partial}{\partial z_i}
\langle p' | \tilde F^{+\rho}(-z/2) \mathcal W F^{+}_{~\, \rho}(z/2) | p \rangle \right|_{z^+ = 0, z_{\perp} = 0} 
\Deq \epsilon^{ij} \Delta_j x\tilde G_g(x)  . \label{tildeG}
\end{align}
After a straightforward calculation with the help of (\ref{der1}), (\ref{tildeG}) implies
\begin{align}
x\tilde G_g(x) &= -2 \int dx'\, \frac{\partial}{\partial x} \tilde N_D(x,x') + 2 \int dx' \, P \frac{1}{x-x'} \Big ( \frac{\partial}{\partial x} - \frac{\partial}{\partial x'} \Big ) \tilde N_F(x,x')  \notag
\\
&= \frac{d}{dx} (x C_g(x) ) + 4 \int dx' \, P \frac{1}{(x-x')^2} \tilde N_F(x,x'),
\label{tildeGg expr}
\end{align}
where we have used  \eqref{gluegpd} in the second equality. Inserting  \eqref{tildeGg expr} into the equations of motion relation \eqref{first} we get the differential equation
\begin{align}
x \frac{d}{dx} C_g(x) - C_g(x)  &= -x \Delta G(x) + 2 G(x) - 4 \int dx' \, P \frac{2x-x'}{x(x-x')^2} \tilde N_F(x,x') \notag
\\
&\quad - \frac{4}{x} \int dx'\, P \frac{1}{x-x'} N_F(x,x') - \frac{4}{x} \sum_q \int dX\, \Psi_{Fq}\left(X+\frac{x}{2},X-\frac{x}{2}\right). \label{deq gluon}
\end{align}
Solving  \eqref{deq gluon} with the boundary conditions $C_g(\pm 1) = 0$ we arrive at 
\beq
C_g(x) &=& x\int_x^{\epsilon(x)}\frac{dx'}{x'}\Delta G(x') -2x\int_x^{\epsilon(x)}\frac{dx'}{x'^2}G(x') + 4x\sum_q\int_x^{\epsilon(x)}\frac{dx'}{x'^3}\int dX \Psi_{Fq}\left(X+\frac{x'}{2},X-\frac{x'}{2}\right) \label{final2}  \nn && +4x\int_x^{\epsilon(x)}dx_1\int dx_2P\frac{N_F(x_1,x_2)}{x_1^3(x_1-x_2)} +4x\int_x^{\epsilon(x)}dx_1\int dx_2\frac{\tilde{N}_F(x_1,x_2)}{x_1^3(x_1-x_2)}P\frac{2x_1-x_2}{x_1-x_2}. 
\eeq
Similarly to the quark case, in Appendix \ref{ab}, we use a different form of the Lorentz invariant relation to  derive the same result.   

 (\ref{final1}) and  (\ref{final2}) are the main results of this paper. As promised, these formulas achieve the decomposition of  the quark and gluon spin-orbit correlations $C_{q,g}(x)$ into the WW part  related to the twist-two PDFs and the genuine twist-three contributions. 
The WW part of (\ref{final2}) agrees with the one reported in \cite{Bhattacharya:2024sck}. 
As pointed out already, $C_g(x)$ is odd in $x$, so the first moment vanishes $\int_{-1}^1 dx C_g(x)=0$. The corresponding results for spin-0 hadrons are trivially obtained by setting $\Delta q,\Delta\bar{q},\Delta G, G_F,\tilde{G}_F,F,h_1$ to zero.

Similarly to the quark case discussed around (\ref{blu}),  we notice a connection between (\ref{final2}) and a known distribution function relevant to a transversely polarized nucleon. That is the ${\cal G}_{3T}(x)$ distribution which is the gluonic counterpart of the $g_T(x)$ distribution \cite{Ji:1992eu,Hatta:2012jm,Benic:2021gya}.
\beq
\int \frac{d\lambda}{2\pi}e^{ix\lambda}\langle p,s_\perp|F^{n\alpha}(0){\cal W}\tilde{F}^i_{\ \alpha}(\lambda n)|p,s_\perp\rangle = 2ix{\cal G}_{3T}(x)s^i,
\eeq
${\cal G}_{3T}$ admits a twist decomposition \cite{Hatta:2012jm} similar to (\ref{gt}) 
\begin{align} \notag
x^2{\cal G}_{3T}(x) &=\tilde{g}_G(x) -\sum_q \int dXG_{Fq}\left(X+\frac{x}{2},X-\frac{x}{2}\right)
\\
&\quad +\int dx'P\frac{F(x,x')-F(x,x-x')-2F(x',x'-x)}{x-x'},
%(F(x,x')-F(x,x-x')-2F(x',x'-x)),
\end{align}
where $F$ is as defined in (\ref{yoshida}) and the kinematical twist-three part $\tilde{g}_G(x)$ is what is called   $\tilde{g}(x)$ in \cite{Hatta:2012jm}.  
Explicitly, (see (3.22) of \cite{Hatta:2012jm})  
\beq
&& \frac{2\tilde{g}_{G}(x)}{x}= x\int_x^{\epsilon(x)}\frac{dx'}{x'}\Delta G(x')+ 2x\sum_q\int_x^{\epsilon(x)}\frac{dx'}{x'^3}\int dX G_{Fq}\left(X+\frac{x'}{2},X-\frac{x'}{2}\right) \\ &&\quad  +2x\int_x^{\epsilon(x)}dx_1\int dx_2P\left[\frac{-2(F(x_1,x_2)-F(x_2,x_2-x_1))}{x_1^3(x_1-x_2)} +\frac{F(x_2,x_2-x_1)-F(x_1,x_1-x_2)}{x_1^2(x_1-x_2)^2}\right]. \nonumber
\eeq
Using this, we can rewrite  (\ref{final2}) as 
\beq 
&&C_g(x)=\frac{2\tilde{g}_G(x)}{x} -2x\int_x^{\epsilon(x)}\frac{dx'}{x'^2}G(x')+ 4x\sum_q\int_x^{\epsilon(x)}\frac{dx'}{x'^3}\int dX \Lambda_{q}\left(X+\frac{x'}{2},X-\frac{x'}{2}\right) \label{cgxx} \\ &&  
 \quad +4x\int_x^{\epsilon(x)}dx_1\int dx_2P\left[\frac{2(\Lambda_G(x_1,x_2)-\Lambda_G(x_2,x_2-x_1))}{x_1^3(x_1-x_2)} +\frac{\Lambda_G(x_1,x_1-x_2)-\Lambda_G(x_2,x_2-x_1)}{x_1^2(x_1-x_2)^2}\right],  \nonumber 
\eeq
where $\Lambda_G$, introduced in (\ref{yoshida}), is an intrinsically off-forward distribution.   Therefore,  $C_g(x)$ is partially related to the $\tilde{g}_G(x)$ distribution for a transversely polarized nucleon. We however note that  the right hand side of   (\ref{cgxx}) is dominated by the unpolarized gluon distribution \cite{Bhattacharya:2024sno,Bhattacharya:2024sck}. Besides, $\tilde{g}_G=0$ for spinless hadrons.

Returning to (\ref{final2}) we compute the  second moment as 
%\beq
%\int_{-1}^1 dx x C_g(x) &=& \frac{1}{3}\int_{-1}^1dx \Biggl[x^2\Delta G(x) -2xG(x)-4\sum_q\int dX\Psi_{Fq}\left(X,x\right) 
%\nn && +4\int dx'\frac{\tilde{N}_F(x,x')}{x-x'} 
%+6\int dx'\frac{\tilde{N}_F(x,x')}{x-x'}\Biggr].
%\eeq
\beq
C_g^{(2)}&\equiv& \int_0^1dx xC_g(x) \nn &=& \frac{1}{3}\int_0^1dx x^2\Delta G(x) -\frac{2}{3}A_g+\frac{2}{3}\sum_q\int dx dx' \Psi_{Fq}(x,x')+\int_{-1}^1 dx dx'\frac{\tilde{N}_F(x,x')}{x-x'}. \label{cg1}
\eeq
where 
\beq
A_g=\int_0^1dx xG(x),
\eeq
is the momentum fraction of the nucleon carried by gluons. Compare to the quark case (\ref{cq1}). We will discuss more about the second moments in the next section. 
Finally,  the integral of (\ref{first}) leads to another formula  
\beq
\int_{-1}^1dx x^2\tilde{G}_g(x)= \frac{4}{3}A_g-\frac{2}{3}\int_0^1 dx x^2\Delta G(x) -\frac{4}{3}\sum_q\int dx dx' \Psi_{Fq}(x,x'), \label{ggmoment}
\eeq
while $\int dx x \tilde{G}_g(x)=0$ since $x\tilde{G}_g(x)$ is odd in $x$. 
 Compare with (\ref{g2}). We can eliminate $\Psi_F$ from these equations and obtain 
\begin{align} \notag 
&\frac{3}{2} \int_{-1}^1 dx x^2\left(4\sum_q\tilde{G}_q(x)-\tilde{G}_g(x)\right) 
\\
&=4\sum_q A_{q+\bar{q}}-2A_g 
%-4\Delta\Sigma^{(3)}+\Delta G^{(3)} 
-\sum_q\frac{4m_q}{M}H_{1q}^{(2)}.
-\int_0^1 dx x^2(4\Delta \Sigma(x)-\Delta G(x)).
\end{align}
The right hand side is expressed purely  in terms of twist-two quantities.

\section{A momentum sum rule}

A salient feature of our main results (\ref{final1}), (\ref{final2}) is that the spin-orbit correlations $C_{q,g}(x)$ receive a dominant contribution from  the unpolarized PDFs $q(x),G(x)$. As a result,  the standard quark and gluon momentum fractions $A_{q,g}$ show up in the second moments (\ref{cq1}) and (\ref{cg1}). By combining these formulas, we can immediately write down a new momentum sum rule 
\beq
1&=& \sum_q A_{q+\bar{q}}+A_g \label{triv}\\
&=&\Delta \Sigma^{(3)}+\frac{1}{2}\Delta G^{(3)}  
%\int_0^1 dx x^2 \left(\Delta \Sigma(x)+\frac{1}{2}\Delta G(x)\right)  
-3 \sum_q C^{(2)}_{q}-\frac{3}{2}C^{(2)}_g  \nn 
&&   +\frac{3}{2}\int_{-1}^1 dx dx'\left[2x \sum_q \frac{ \Psi_{Fq}(x,x')+\tilde{\Psi}_{Fq}(x,x')}{x-x'} +\frac{\tilde{N}_F(x,x')}{x-x'}\right]  +\sum_q\frac{m_q}{M}H_{1q}^{(2)},
\label{main}
\eeq
where we abbreviated 
\beq
\Delta \Sigma^{(3)}\equiv \int_0^1 dx x^2\Delta \Sigma(x) = \sum_q \int_0^1 dx x^2(\Delta q(x)+\Delta \bar{q}(x)), \qquad \Delta G^{(3)}\equiv\int_0^1dx x^2\Delta G(x). \label{secondm}
\eeq 
  Note that the twist-three term involving $\Psi_F$ reduces to the local matrix element (\ref{force})  
  \beq
  \int dx dx' \frac{2x}{x-x'}\Psi_F(x,x') = \int dx dx' \Psi_F(x,x'), \label{reduce}
  \eeq
  where we used  (\ref{psisym}). 
   The other two terms, involving $\tilde{\Psi}_F$ and $\tilde{N}_F$ do so only in the light-cone gauge.
% \JS{How? Are $\tilde{\Psi}_F,\tilde{N}_F$ not gauge invariant?} {\color{red} In the LC gauge, $F^{+i}=\partial_-A^i$. Integration by parts in $\partial_-$ brings down a factor of $x-x'$, canceling the $1/(x-x')$ in the denominator. Integration over $x,x'$ then  makes the matrix element local. You can elaborate on this, if you want. } \JS{I think I understand. So $\tilde{\Psi}_F,\tilde{N}_F$ are gauge invariant and in the light-cone gauge they can be written in terms of a local matrix element while in a general gauge they can not be written as a local matrix element, since it involves the integral $\int_0^{\pm \infty} d\tau \, \tilde{F}^{ni}(\tau n)$. You have shown this in \eqref{mat1} and \eqref{mat2}. Looking at these expressions it is clear that this integral reduces to $A^i(0)$ in the light-cone gauge.} 

Let us make a comparison  with the spin sum rules (\ref{jm}), (\ref{ji})  which can be rewritten in the form
\beq
\frac{1}{2}&=&\frac{1}{2}\sum_q \left(A_{q+\bar{q}}+B_{q+\bar{q}}\right) + \frac{1}{2}(A_g+B_g)
\label{ji2}\\
&=& \frac{1}{2}\Delta\Sigma^{(1)}+\Delta G^{(1)}+L_q^{(1)}+L_g^{(1)}, \label{jm2}
\eeq
%{\color{red} Here it is ok to use the (1) superscripts.}
where 
\begin{align}
   B_{q,g}  \equiv \int_{-1}^1 dx \, x E_{q,g}(x)
\end{align}
are the second moments of the GPDs $E_{q,g}$, which obey the sum rule $\sum_q B_{q+\bar q} + B_g = 0$.
 We remark that $J_{q,g}=\frac{1}{2}(A_{q,g}+B_{q,g})$ are the quark and gluon angular momenta in the Ji sum rule (\ref{ji}). The correspondence between (\ref{triv}), (\ref{main}) and (\ref{ji2}), (\ref{jm2}) is obvious.
 %; they can be derived from exact twist-decomposition formulas for the $C_{q,g}$, \eqref{final1} and \eqref{final2}, and $L_{q,g}$, (2.31) and (3.26) in \cite{Hatta:2012cs}, respectively, using the momentum sum rules for $A_{q,\bar q, g}$ (and $B_{q,\bar q, g}$). 
 %In accordance with the existence of two well-known spin sum rules, we have derived the second momentum sum rule (\ref{main}). 
 Our derivation of (\ref{main}) based on the exact twist decomposition of $C_{q,g}(x)$ closely parallels the derivation of the JM sum rule (\ref{jm2}) from the twist decomposition of $L_{q,g}(x)$   \cite{Hatta:2012cs}. We therefore view  (\ref{main}) as the momentum version of the JM spin sum rule. 
However, unlike the latter, the genuine twist-three distributions explicitly remain in (\ref{main}). (\ref{jm2}) does not contain such terms because they cancel between the quark and gluon sectors (in the derivation of \cite{Hatta:2012cs}). In the present case, the cancellation is incomplete partly because the second moments (rather than the first) of $C_{q,g}(x)$ are involved. 

A key question regarding the new sum rule (\ref{main}) is that the physical meaning of each term  is not immediately obvious. In contrast, all the terms in the JM sum rule have a clear physical meaning. While we do not have a fully satisfactory answer to this question, let us nevertheless  speculate on how the  connection  to `momentum' could be made. For simplicity, we assume a spin-0 hadron  in the chiral limit in which case the sum rule (\ref{main}) takes the simpler form 
\beq
1&=& 
-3 \sum_q C^{(2)}_{q}-\frac{3}{2}C^{(2)}_g    +\frac{3}{2}\int_{-1}^1 dx dx'\left[2x \sum_q \frac{\Lambda_{q}(x,x') +\tilde{\Lambda}_{q}(x,x') }{x-x'}+\frac{\tilde{\Lambda}_G(x,x')}{x-x'}\right],  \label{sum0}
\eeq
where
\beq
\tilde{\Lambda}_G(x_1,x_2)\equiv \Lambda_G(x_1,x_1-x_2)-\Lambda_G(x_2,x_2-x_1).
\eeq
Note that the higher moments of the helicity distributions $\Delta\Sigma^{(3)},\Delta G^{(3)}$ in  (\ref{main}) are strongly suppressed due to the factor $x^2$ in (\ref{secondm}). 
%since the corresponding $x$-distributions are peaked at values $x < 1$. 
Hence (\ref{sum0}) may be a good approximation even for spin-$\frac{1}{2}$ hadrons.  

%In the following, we  ignore prefactors such as  $\frac{3}{2}$ and focus only on the signs. 
%First of all, just like the JM sum rule, the new sum rule can be most naturally interpreted in the infinite momentum frame where the hadron momentum equals the hadron energy. Thus the momentum sum rule is equivalent to the energy sum rule.  
%\JS{I do not understand the above paragraph.}
%{\color{red} Below we identify `potential`, and I normally think that potential is a part of energy or hamiltonian, rather than momentum. This is why I wanted to switch to the language of energy. But I agree, it is not very clear...} \JS{I still do not like the paragraph. Is it necessary to say that?}

First, the negative sign in front of $C_{q,g}^{(2)}$   strongly indicates that  $C^{(2)}_{q,g}$, hence also  the distributions $C_{q,g}(x)$ (for the most part) are negative.  This means that the helicity and OAM of individual partons are anti-aligned. There are already several indications of this in the literature  \cite{Lorce:2014mxa,Tan:2021osk,Engelhardt:2021kdo}, but  most strikingly, when $x\ll 1$  one finds that $C_{q,g}(x)$ are `maximally negative' \cite{Bhattacharya:2024sno} as already mentioned in (\ref{fir}). Intuitively, the approximate proportionality relation at small $x$ (\ref{fir}) can be understood as follows \cite{Bhattacharya:2024sno}. For each quark or gluon, the product $L_3S_3$ (see the introduction) is a negative number of order unity. Roughly, the total $C_{q,g}(x)$ is this number times the number of quarks $q(x)$ or gluons $G(x)$. Therefore,  $C_{q,g}^{(2)} = \int dx xC_{q,g}(x)$ are indeed a measure of momentum fractions $\int dx xq(x),\int dx xG(x)$, up to a negative prefactor. 

% \JS{If I take this line of reasoning literally, I obtain
% \begin{align}
% 1 = -3 \sum_q C_q^{(2)} - \frac{3}{2} C_g^{(2)} + \text{twist-3} ``\simeq" - 3 (-1/2) \sum_q \int dx x q(x) - \frac{3}{2} (-1) \int dx xG(x) = 3/2.
% \end{align}
% Can difference of factor $3/2$ be also understood intuitively?
% } {\color{red} The coefficient $-1/2,-1$ is valid only at small-$x$. For generic $x$, we need to use (6.11). Then you get the correct coefficient of unity.}

In practice, the small-$x$ region does not significantly contribute  to the $x$-integral. Also, the ratios $C_{q}(x)/q(x)$, $C_{g}(x)/G(x)$ will not be  constant when $x\sim 1$. A reasonable approximation for generic values of $x$ is obtained by dropping the twist-three terms in \eqref{final1} and \eqref{final2}, giving
\beq
C_q(x) \approx -x\int_x^{\epsilon(x)} \frac{dx'}{x'^2}q(x'), \qquad C_g(x) \approx -2x\int_x^{\epsilon(x)}\frac{dx'}{x'^2}G(x') ,\label{corr}
\eeq
and this already saturates the sum rule 
\beq
-3\sum_q C_q^{(2)}-\frac{3}{2} C_g^{(2)}\approx \sum_q A_{q+\bar{q}}+A_g=1. \label{sat}
\eeq
Using this as motivation, we interpret the $C_{q,g}^{(2)}$ terms in (\ref{sum0}) as contributions due to the quark and gluon `kinetic energy' 
\beq
-3C_q^{(2)}\equiv T_q, \qquad -\frac{3}{2}C_g^{(2)}\equiv T_G.
\eeq

The corrections to (\ref{sat}) are accounted for by the  twist-three terms in (\ref{sum0}). Among them,  the $\Lambda_q$ term is special in that it reduces to the local matrix element (\ref{force}). (See (\ref{reduce}), $\Psi_F=\Lambda$ for a spin-0 hadron.)     According to \cite{Aslan:2019jis} (see also \cite{Crawford:2024wzx}),  the physical interpretation of (\ref{force}) is that, when Fourier transformed to impact parameter ($b_\perp$) space,  it represents  the transverse `color Lorentz force' ${\cal F}_\perp$ acting on quarks. Indeed, the operator $F^{+i}=F^{+i}_at^a$ in (\ref{force}) can be written in terms of the color electromagnetic fields 
\beq
F_a^{+i}=\frac{1}{\sqrt{2}}(\vec{E}+\vec{v}\times \vec{B})_a^i,
\eeq
acting on the quark color electric charge density $g\bar{q}\gamma^+t_aq$. The velocity $\vec{v}=(0,0,-1)$ represents the trajectory of a quark when it is probed in physical processes. (Imagine a quark struck by a virtual photon in Deep Inelastic Scattering.) Following \cite{Aslan:2019jis}, we write 
\beq
\frac{3i\Delta^i}{2}\int \frac{ d^2\Delta_\perp}{(2\pi)^2} e^{-ib_\perp \cdot \Delta_\perp } \int dx dx'\Lambda_q(x,x',\Delta_\perp^2) ={\cal F}_q^i(b_\perp)\equiv -\frac{\partial}{\partial b^i} V_q(b_\perp), \label{burk}
\eeq
where we temporarily reinstated the $\Delta_\perp$-dependence in $\Lambda_q$ and adopted the same sign convention of ${\cal F}^i$ as in \cite{Aslan:2019jis}.\footnote{The sign convention of the QCD coupling $g$ in \cite{Aslan:2019jis} is opposite from ours. } We however introduced a different normalization factor which is suggested by the sum rule (\ref{sum0}). In the last equality, we have introduced the corresponding `potential'  $V_q(b_\perp)$ whose gradient gives the Lorentz force. (\ref{burk}) can be solved as  
\beq
\frac{3}{2}\int dxdx' \Lambda_q(x,x',\Delta_\perp=0) = \int d^2b_\perp V_q(b_\perp)\equiv V_q.  \label{poten}
\eeq
Therefore, the $\Lambda_q$ term in (\ref{sum0}) can be interpreted as the  potential energy $V_q$ associated with the color Lorentz force.

The interpretation of the $\tilde{\Lambda},\tilde{\Lambda}_G$ terms are more involved since they do not reduce to the matrix element of a local operator  (except in the light-cone gauge). In fact, a similar operator appears already in the first moment, as the difference between $C_q$ (\ref{firstm}) and $C_q^{\rm kin}$ (\ref{dlo}) 
\beq
 \langle p'|\bar{q}(0)\Slash n  \gamma_5 gt^b q(0) \int_0^{\pm \infty}d\tau \mathcal W^{ba}_{0, \tau n}\tilde{F}_a^{ni}(\tau n) |p\rangle \Deq i\Delta^i \int dx dx' \frac{\tilde{\Lambda}_q(x,x')}{x-x'} ,  \label{difference}
\eeq
cf. (\ref{qgq2}), 
where the $\pm$ sign in the integration limit does not matter since $\tilde{\Lambda}_q$ vanish at $x=x'$. 
%\JS{Actually, if since we use the principal value prescription corresponds to the average between $+$ and $-$, so I am not sure if saying ``it depends on the boundary conditions is actually correct". However, I agree that this does not matter here, since $\tilde{\Lambda}_q(x,x) = 0$.} {\color{red} I don't want to explicitly write the average since it's awkward.}
The physical interpretation of (\ref{difference}) is as follows. 
 The dual field strength tensor has the components 
\beq
\tilde{F}_a^{+i}=-\frac{1}{\sqrt{2}}(\vec{B}-\vec{v}\times \vec{E})_a^i. \label{feel}
\eeq
In electrodynamics, this is known as the would-be Lorentz force acting on a magnetic  monopole \cite{griffiths}. Although a quark does not carry a magnetic charge,  the nonvanishing matrix element (\ref{difference}) shows that it can nonlocally couple to the transverse force (\ref{feel}). Classically, this shifts the quarks transverse position  and thereby changes its orbital angular momentum $b_\perp \times k_\perp$.
%\JS{About the statement: ``which shifts its transverse position $b_\perp$ and thereby changes its orbital angular momentum [to] $b_\perp \times k_\perp$". I guess this is by itself a classical E\&M statement and you are carrying over the QCD case to appeal to intuition? (I just want to understand the reasoning)} {\color{red} yes}
Since the sign of this force is helicity-dependent as indicated by the presence of a $\gamma_5$, the interaction generates an extra amount of helicity-orbit correlation which contributes to the difference $C_q-C_q^{\rm kin}$.   This is similar to the explanation of the potential OAM $L_q-L_q^{\rm kin}$ in terms of `torque' a quark gets during the initial or final state interaction \cite{Burkardt:2012sd}. 

Returning to (\ref{sum0}), we can write the last two terms as 
\beq
\langle p'|\bar{q}(0) \Slash n\gamma_5g\left(t^bi\overrightarrow{D}^n -i\overleftarrow{D}^nt^b\right) q(0) \int_0^{\pm \infty}d\tau  \mathcal W^{ba}_{0, \tau n}\tilde{F}_a^{ni}(\tau n) |p\rangle \Deq i\Delta^i\int dx dx' \frac{2x\tilde{\Lambda}_q(x,x')}{x-x'} , \label{mat1}
\eeq
\beq
\langle p'| i\tilde{F}^{n\rho}(0)gT^b F^n_{\ \rho}(0)\int_0^{\pm \infty}d\tau \mathcal W^{ba}_{0, \tau n}\tilde{F}_a^{ni}(\tau n)|p\rangle \Deq i\Delta^i \int dx dx' \frac{\tilde{\Lambda}_G(x,x')}{x-x'}. \label{mat2}
\eeq 
%\beq
%\int \frac{d\tau}{2\pi} \int d(x-x') e^{i\tau(x'-x)}\frac{i\tilde{F}^{+i}(\tau n)}{x-x'} = \int_0^{\pm \infty}d\tau \tilde{F}^{+i}(\tau n), \label{mat}
%\eeq
The composite quark operator in (\ref{mat1}) is proportional to the `time' derivative $\partial^+$ of the axial color charge operator $\bar{q}\gamma^+\gamma_5gt^a q$ in (\ref{difference}). The gluonic operator in (\ref{mat2}) may seem unfamiliar. To get a rough physical intuition, consider the Bianchi identity 
\beq
0=D_\rho \tilde{F}^{\rho +} = \partial_\rho \tilde{F}^{\rho +} +ig[A_\rho, \tilde{F}^{\rho+}].
\eeq
This can be written as 
\beq
\partial_\rho \tilde{F}_a^{\rho +} = -i\tilde{F}^{+\rho}gT^a A_\rho,
\eeq
which may be viewed as a relativistic analog of the field equation for a magnetic monopole $\vec{\nabla} \cdot \vec{B} =\rho_m$ where the operator $-i\tilde{F}^{+\rho}gT^a A_\rho$ behaves as an effective magnetic charge density. This is similar to the operator in (\ref{mat2}) except that the latter involves an extra $z^-$ derivative, $F^+_{\ \rho}=\partial^+ A_\rho$ (in the light-cone gauge) related to the energy weight factor $x$.

 As in (\ref{burk}), (\ref{poten}), we introduce the corresponding potentials
 \beq
\frac{3}{2}\int dx dx' \frac{2x\tilde{\Lambda}_q(x,x')}{x-x'}=\int d^2b_\perp \tilde{V}_q(b_\perp)=\tilde{V}_q,\nn
 \frac{3}{2}\int dx dx' \frac{\tilde{\Lambda}_G(x,x')}{x-x'} =\int d^2b_\perp \tilde{V}_G(b_\perp)=
 \tilde{V}_G,
 \eeq
 that source the `dual' Lorentz force. 
Therefore, the last two terms in (\ref{sum0}) can also be interpreted as potential energy contributions.   
  Finally, it is worthwhile to mention that the presence of the $\Lambda_q$ term in (\ref{sum0}) actually requires the  $\tilde{\Lambda}_q$ term. Indeed, they can be combined as 
\beq 
 \int dx  dx' 2x\frac{\Lambda_q(x,x')+\tilde{\Lambda}_q(x,x')}{x-x'}= \int dx  dx' x \frac{\Lambda_q(x,x')+\tilde{\Lambda}_q(x,x')+\Lambda_q(-x,-x')+\tilde{\Lambda}_q(-x,-x')}{x-x'}.\nn
\eeq
The linear combination of the four terms is charge parity ($C$) even ($PT$-even)  \cite{Braun:2009mi,Hatta:2019csj},\footnote{Compare with (27) of \cite{Hatta:2019csj}. The sign difference is because $L_q(-x)=L_{\bar{q}}(x)$ for the quark OAM but $C_q(-x)=-C_{\bar{q}}(x)$ for the quark spin-orbit coupling.} and is associated with the operator 
\beq
\bar{q}(-\lambda n/2)W\gamma^+g\left(F^{+i}+i\gamma_5\tilde{F}^{+i}\right)(\mu n)Wq(\lambda n/2).
\eeq 
They must appear in this particular combination in $C$-even observables such as the total momentum.

To summarize, the sum rule (\ref{sum0}) physically represents the decomposition of the total hadron energy into the kinetic and potential energies 
\beq
1=\sum_q T_q+T_G+\sum_q V_q+\sum_q \tilde{V}_q+\tilde{V}_G,
\eeq
where the latter are associated with the color electric and magnetic Lorentz forces. 
Admittedly, the above argument is only qualitative and the extension to the spin-$\frac{1}{2}$ case is nontrivial. Nevertheless, we believe our interpretation offers a novel perspective on hadron momentum/energy that deserves further investigation, especially because  (\ref{main}) or (\ref{sum0}) is exact.  At least, since  the new sum rule  involves   at most twist-three distributions, it can be tested in experiments in the same sense that the JM sum rule can be tested.

\section{Conclusions}

In this paper, we have derived exact new formulas (\ref{final1}), (\ref{final2}) for the  quark and gluon spin-orbit correlations $C_{q,g}(x)$  which completely reveal their twist structure. These results are  applicable to both spin-$\frac{1}{2}$ and spin-0 hadrons. The only difference is that, in the former case,  the kinematical twit-three distributions $\tilde{g}$ and $\tilde{g}_G$, familiar in the context of transverse spin physics, explicitly enter the formulas as shown in  (\ref{blu}) and (\ref{cgxx}).  Based on these results, we have derived a new momentum  sum rule (\ref{main}) which is the momentum version of the JM sum rule. We have discussed its physical interpretation focusing on the spin-0 case (\ref{sum0}) and pointed out the connection to the color Lorentz force. However, more work is needed to fully grasp  the physics content of the sum rule.

 As stated in the introduction, our work is partly motivated by a recent observation regarding the peculiar behavior (\ref{fir}) of $C_{q,g}(x)$ at small-$x$ \cite{Bhattacharya:2024sno}. Let us quickly check the consistency with our results. From (\ref{final1}) and (\ref{final2}), one can immediately deduce that the small-$x$ limit of $C_{q,g}(x)$ is dominated by the unpolarized PDFs 
 %\JS{[Need to clarify and be consistent with use of $\approx, \sim, \simeq, \propto$ in the following. Is $\approx$ and $\sim$ the same here? Maybe better to use $\simeq$?]}
\begin{equation}
\begin{split}
& C_q(x) \approx C_{\bar{q}}(x)\approx  -x\int_x^1 \frac{dx'}{x'^2}q(x') \approx -\frac{1}{2+c}q(x), \\
& C_g(x) \approx -2x\int_x^1 \frac{dx'}{x'^2}G(x') \approx -\frac{2}{2+c}G(x), \label{yu}
\end{split}
\end{equation}
where in the last expression we assumed a Regge behavior $q(x)\sim G(x) \sim 1/x^{1+c}$ ($0<c<1$). If $c\ll 1$ as suggested by perturbation theory $c = \mathcal O( \alpha_s )$, this gives $C_q(x)\approx -\frac{1}{2}q(x)$ and $C_g(x)\approx -G(x)$, in agreement with (\ref{fir}). The correction to (\ref{yu}) from the polarized PDFs are strongly suppressed at small-$x$. Indeed, 
\beq
x\int_x^1 \frac{dx'}{x'}\Delta q(x') \sim x\int_x^1 \frac{dx'}{x'}\Delta G(x') \sim x^{1-d}\to 0,
\eeq
where $\Delta q(x)\sim \Delta G(x)\sim 1/x^d$ ($0<d<1$). It remains to be seen whether the genuine twist-three contributions  affect the leading behavior (\ref{yu}). This can be tested by adapting the theoretical frameworks developed to study the OAM distributions $L_{q,g}(x)$ at small-$x$ \cite{Kovchegov:2023yzd,Manley:2024pcl}. 

Now that the complete twist decomposition of $C_{q,g}(x)$ is available, their renormalization group scale dependence can be studied following a strategy used  for the OAM distributions $L_{q,g}(x)$ \cite{Hatta:2019csj}. This is based on the observation that, at zero skewness, the evolution of twist-three, off-forward correlators is identical to that of forward correlators for which the evolution equation is known \cite{Braun:2009mi}. Such an analysis will be important in the future as the phenomenology of spin-orbit correlation becomes more developed.

Last but not least, it is of course very important to experimentally test the new momentum sum rule (\ref{main}) or (\ref{sum0}). In a sense, this is the whole point of this type of sum rule. Despite its apparent complexity, the situation might be better than the experimental measure of the parton OAMs $L_{q,g}$ in the JM sum rule since the polarization of the target hadron is not required to access the spin-orbit correlation.  Several observables have been already proposed in the literature \cite{Bhattacharya:2017bvs,Boussarie:2018zwg,Bhattacharya:2023hbq,Bhattacharya:2024sck}. 
We hope our sum rule further motivates  experimental activities, especially at the future Electron-Ion Collider.

\section*{Acknowledgements}

We thank Vladimir Braun and Shinsuke Yoshida   for correspondence. 
This work was supported by the U.S. Department of Energy under Contract No. DE-SC0012704, and also by  Laboratory Directed Research and Development (LDRD) funds from Brookhaven Science Associates.

\appendix 

\section{Useful formulas}
\label{aa}

Here we list some useful formulas which will help to follow the derivation of various results in the main body of the paper. We use the short-hand notation for a straight-Wilson line along the light-cone some transverse position $z_{\perp}$
\beq
W_{\tau\lambda}(z_\perp) \equiv W_{\tau n + z_{\perp}, \lambda n + z_{\perp}} = P\exp\left(-ig\int_{\lambda}^{\tau} d\tau' \, n \cdot A(\tau' n,z_\perp)\right).
\eeq
We have the following identity
\begin{align} \notag
&D^i(\tau n + z_{\perp})W_{\tau \lambda}(z_{\perp}) - W_{\tau\lambda}(z_{\perp})D^i(\lambda n+ z_{\perp})
\\
&\quad = i\int_{\lambda}^{\tau}d\tau' W_{\tau\tau'}(z_{\perp})gF^{ni}(\tau'n+ z_{\perp})W_{\tau'\lambda}(z_{\perp}),
\label{WDid1}
\\ \notag
&W_{\tau\lambda}(z_{\perp}) \overleftarrow{D}^i(\lambda n + z_{\perp}) - \overleftarrow{D}^i(\tau n + z_{\perp})W_{\tau\lambda}(z_{\perp})
\\
&\quad = i\int_{\lambda}^{\tau} d\tau' W_{\tau\tau'}(z_{\perp})gF^{ni}(\tau' n + z_{\perp})W_{\tau'\lambda}(z_{\perp}),
\label{WDid2}
\end{align}
which should be viewed as an identity of differential operators acting on the right and left on quark fields depending on $z_{\perp}$. There are completely analogous relations for the adjoint representation acting on gluons field strength tensors depending on $z_{\perp}$.
Adding \eqref{WDid1} and \eqref{WDid2} gives
\begin{align} \notag
&\overleftrightarrow{D}^i(\tau n + z_{\perp})W_{\tau\lambda}(z_{\perp}) - W_{\tau\lambda}(z_{\perp})\overleftrightarrow{D}^i(\lambda n + z_{\perp})
\\
&= + i\int_{\lambda}^{\tau} d\tau' W_{\tau\tau'}(z_{\perp}) gF^{ni}(\tau' n + z_{\perp})W_{\tau' \lambda}(z_{\perp}).
\end{align}
To obtain (\ref{delta}) and (\ref{cgdef}), the following formula is useful
\beq
\int d\tau e^{i\tau (x_2-x_1)} \int^\tau_\lambda d\tau' &=& \int d\tau e^{i\tau(x_2-x_1)} \int_{-\infty}^\infty d\tau' (\theta(\tau-\tau')\theta(\tau'-\lambda)-\theta(\lambda-\tau')\theta(\tau'-\tau)) \nn 
&=&i\int d\tau' e^{i\tau'(x_2-x_1)}\left(\frac{\theta(\tau'-\lambda)}{x_2-x_1+i\epsilon}-\frac{\theta(\lambda-\tau')}{x_1-x_2+i\epsilon}\right) \nn 
&=& i\int d\tau' e^{i\tau'(x_2-x_1)}\left(P\frac{1}{x_2-x_1}-i\pi \delta(x_1-x_2)\epsilon(\tau'-\lambda)\right).
\eeq

Next consider the derivative of the nonlocal operator product \cite{Balitsky:1987bk}
\beq
\frac{\partial}{\partial z^\mu} \bar{q}(-uz)W_{-uz,uz}\gamma^\alpha q(uz) &=&  u\bar{q}(-uz)\left(W_{-uz,uz} D_\mu(uz) -\overleftarrow{D}_\mu(-uz)W_{-uz,uz}\right)\gamma^\alpha q(uz) \nn && + i \int_{-u}^u d\tau  \tau \bar{q}(-uz)W_{-uz,\tau z}gF_{\mu\nu}(\tau z) x^\nu W_{\tau z,uz}\gamma^\alpha q(uz). \label{der1}
\eeq
This formula is valid for generic $z^\mu$, not necessarily on the light-cone. On the other hand, the translation derivative (see (\ref{trans})) of the same operator reads  
\beq
  {\mathfrak D}_\mu\Bigl[ \bar{q}(-uz)W_{-uz,uz}\gamma^\alpha q(uz) \Bigr] &=& \bar{q}(-uz)\left(W_{-uz,uz}D_\mu(uz) +\overleftarrow{D}_\mu(-uz)W_{-uz,uz}\right)\gamma^\alpha q(uz) \nn && + i \int_{-u}^u d\tau \bar{q}(-uz)W_{-uz,\tau z}gF_{\mu\nu}(\tau z) z^\nu W_{\tau z,uz}\gamma^\alpha q(uz). \label{der2}
\eeq
Note the extra factor of $\tau$ in the integrand of (\ref{der1}) which is absent in (\ref{der2}). 
In deriving \eqref{tildeGq expr}, this factor is replaced by (see \eqref{inverse})
\beq
\tau e^{-i\frac{\lambda}{2}(x_1+x_2)-i\tau(x_2-x_1)} = \frac{-i}{2}\left(\frac{\partial}{\partial x_1}-\frac{\partial}{\partial x_2}\right) e^{-i\frac{\lambda}{2}(x_1+x_2)-i\tau(x_2-x_1)},
\label{partial}
\eeq
and then partial integral is utilized. 

\section{Alternative derivation of (\ref{final1}) and (\ref{final2})  }
\label{ab}

In this appendix, we give an alternative derivation of our main results (\ref{final1}) and (\ref{final2}) which is closely related to the method used in \cite{Hatta:2012cs}. 

\subsection{Quark case}

Let us define \cite{Balitsky:1987bk}
\beq
I&\equiv&\left. z^\mu \left(\frac{\partial}{\partial z^\mu} \langle p'|\bar{q}(-z/2)\gamma^i\gamma_5 Wq(z/2)|p\rangle -\frac{\partial}{\partial z_i}\langle p'|\bar{q}(-z/2)\gamma_\mu\gamma_5 Wq(z/2)|p\rangle \right) \right|_{z^+ = 0, z_{\perp} = 0}.
\eeq
We shall  evaluate this  in two ways. First, inserting (\ref{gen}), we get 
\beq
I&=& \int dx e^{-ixP^+z^-} \left[-\frac{d}{dx}(x(\tilde{H}_q+\tilde{G}_q)) +\tilde{G}_q)\right] \bar{u}\gamma^i\gamma_5u  \nn 
&\Deq& -\int dx e^{-ixP^+z^-} \left[\frac{d}{dx}(x\Delta q) +x\frac{d}{dx}\tilde{G}_q\right] i\epsilon^{ij}\Delta_j.  \label{off}
\eeq
On the other hand, by explicitly performing the derivatives using the formulas in Appendix \ref{aa} and then the Dirac equation, we find  
\beq
I
%&=& -\frac{i}{2}z^- \epsilon^{i+}_{\ \ \nu\rho}  {\mathfrak D}^\nu \left(\bar{q}(-z/2)\gamma^\rho Wq(z/2)\right) \nn 
%&&-\frac{z^-}{2}\int_{-\frac{1}{2}}^{\frac{1}{2}}du \epsilon^{i+}_{\ \ \nu\rho}\bar{q}(-z/2)\gamma^\rho WgF^{\nu\tau}(uz)z_\tau Wq(z/2)\nn
%&& -iz^\mu \int^{\frac{1}{2}}_{-\frac{1}{2}} du u \bar{q}(-z/2)\gamma_\mu \gamma_5WgF^{i\tau}(uz)z_\tau Wq(z/2) \nn 
&=&  \frac{z^-}{2} \epsilon^{ij}\Delta_{j}\langle p'|\bar{q}(-z^-/2)\gamma^+ Wq(z^-/2)|p\rangle -im_q\epsilon^{ij}z^- \langle p'|\bar{q}(-z^-/2)\sigma^+_{\ j}Wq(z^-/2)|p\rangle \nn 
&& +\frac{(z^-)^2}{2}\epsilon^{ij}\int_{-\frac{1}{2}}^{\frac{1}{2}}du \langle p'|\bar{q}(-z^-/2)\gamma^+ WgF^+_{\ \ j}(uz^-) Wq(z^-/2)|p\rangle\nn
&&  +i(z^-)^2 \int^{\frac{1}{2}}_{-\frac{1}{2}} du\, u \langle p'|\bar{q}(-z^-/2)\gamma^+ \gamma_5WgF^{+i}(uz^-) Wq(z^-/2)|p\rangle .
\eeq
%The first term on the right hand side can be written in terms of the unpolarized quark PDF $q(x)$ 
%\beq
%-i\epsilon^{ij}\Delta_j\int dx e^{-ixP^+z^-} \frac{d}{dx} q(x).
%\eeq
By equating the two results, we obtain 
\beq
\frac{d}{dx}(x\Delta q) +x\frac{d}{dx}\tilde{G}_q(x)  &=& \frac{d}{dx}q(x)  - \int dx'P\frac{1}{x-x'}\left(\frac{\partial}{\partial x}-\frac{\partial}{\partial x'}\right) \tilde{\Psi}_{Fq}(x,x') \nn &&-\int dx' P\frac{1}{x-x'}\left(\frac{\partial}{\partial x}+\frac{\partial}{\partial x'}\right)\Psi_{Fq}(x,x') -\frac{m_q}{M}\frac{d}{dx}H_{1q}(x).
\eeq
Solving this equation for $\tilde{G}_q$ with the boundary condition $\tilde{G}_q(\pm 1)=0$ and inserting the solution into (\ref{find}), we arrive at   (\ref{final1}).

\subsection{Gluon case }

 Consider the following matrix element
\beq
K&\equiv&  z^\mu \Biggl(\frac{\partial}{\partial z^\mu} z_\beta \langle p'|\tilde{F}^{i\rho}(-z/2){\cal W}F^\beta_{\ \rho}(z/2)+\tilde{F}^{\beta\rho}(-z/2){\cal W}F^{i}_{\ \rho}(z/2)|p\rangle \nn 
&& \qquad \left. -\frac{\partial}{\partial z_i} z_\beta \langle p'|\tilde{F}_\mu^{\ \rho}{\cal W}F^\beta_{\ \rho}+\tilde{F}^{\beta\rho}{\cal W}F_{\mu\rho}|p\rangle  \Biggr)\right|_{z^+ = 0, z_{\perp} = 0}, 
\eeq 
where $z^\mu$ is off the light-cone before taking the derivative. We shall evaluate this in two ways. First, we generalize (\ref{gluegpd}) to  off the light-cone and contract with $z_\beta$ using $\gamma_\perp^\beta = \gamma^\beta -\frac{\gamma\cdot z}{P\cdot z}P^\beta$ 
\beq
&&  z_\beta\langle p'|\tilde{F}^{\alpha\rho}(-z/2){\cal W}F^\beta_{\ \rho}(z/2)+\tilde{F}^{\beta\rho}{\cal W}F^{\alpha}_{\ \rho}|p\rangle  \nn
&& = -\frac{i}{2}\int dx e^{-ixP\cdot z}\bar{u}(p')\left((P\cdot z\gamma^\alpha + P^\alpha z\cdot \gamma)(\tilde{H}_g+\tilde{G}_g)-2P^\alpha \gamma\cdot z \tilde{G}_g\right)\gamma_5u(p). 
\eeq
From this it immediately follows that 
\beq
K\Deq -\frac{P^+z^-}{2}\int dx e^{-ixP^+z^-}\left\{\frac{d}{dx}(x\tilde{H}_g(x))+x^2\frac{d}{dx}\frac{\tilde{G}_g}{x}\right\}\epsilon^{ij}\Delta_j.
\label{offglue}
\eeq
 On the other hand, by performing the derivatives we get 
\beq
K&=&\left. (z^-)^2 \langle p'|\left( -(D^+\tilde{F}^{i\rho}){\cal W} F^+_{\  \rho}+\tilde{F}^{+\rho}{\cal W}(D^+F^{i}_{\ \rho})  -2\frac{\partial}{\partial z_i}  \tilde{F}^{+\rho}{\cal W}F^+_{\ \rho}  \right)|p\rangle \right|_{z^+ = 0, z_{\perp} = 0}. \label{5}
\eeq 
The first two terms can be written as 
\beq
-(D^+\tilde{F}^{i\rho}){\cal W}F^+_{\  \rho}+\tilde{F}^{+\rho}{\cal W}(D^+F^{i}_{\ \rho}) &=& (D^i\tilde{F}^{\rho+}+D^\rho \tilde{F}^{+i}-\epsilon^{+i\rho\sigma}D^\alpha F_{\alpha\sigma}){\cal W}F^+_{\ \rho} \nn 
&& -\tilde{F}^{+\rho}{\cal W}(D^iF_\rho^{\ +}+D_\rho F^{+i}) .
\label{6}
\eeq
 The $D^i$ terms in (\ref{6}) cancel with those from the last term in (\ref{5}) such that 
 \beq
K 
&=& (z^-)^2\langle p'|\Bigl( D^\rho \tilde{F}^{+i}{\cal W}F^+_{\ \rho}-\tilde{F}^{+\rho}{\cal W}D_\rho F^{+i} + \epsilon^{ij}D_\alpha F^{\alpha+}{\cal W}F^{+}_{\ \, j}  \nn 
 && \qquad -2iz^- \int_{-\frac{1}{2}}^{\frac{1}{2}} d\tau \tau \tilde{F}^{+\rho}{\cal W} gF^{i+}(\tau z^-){\cal W}F^+_{\ \ \rho} \Bigr)|p\rangle.
 \eeq
In the first two terms we can write 
\beq
 D^\rho \tilde{F}^{+i}F^+_{\ \rho}-\tilde{F}^{+\rho}D_\rho F^{+i}
&=&    {\mathfrak D}_\rho(\tilde{F}^{+i}F^{+\rho}-\tilde{F}^{+\rho} F^{+i}) +  \tilde{F}^{+i}D_\rho F^{\rho +} \nn && +iz^-\int_{-\frac{1}{2}}^{\frac{1}{2}} d\tau \left(\tilde{F}^{+i}gF^+_{\ \ \rho}(\tau z)F^{+\rho} -\tilde{F}^{+\rho}gF^+_{\ \ \rho}F^{+i}\right) ,
\eeq
and again use (\ref{2di}) to obtain
\beq
K&\Deq & -\epsilon^{ij}\Delta_jP^+z^- \int dx e^{-ixP^+z^-} \Biggl[ \frac{d}{dx}\left(xG(x)  -2\sum_q\int dX\Psi_{qF}\left(X+\frac{x}{2},X-\frac{x}{2}\right)\right) \nn
&& -2\int P\frac{dx'}{x-x'} \left(\frac{\partial}{dx}+\frac{\partial}{\partial x'}\right)N_F(x,x') -2\int \frac{dx'}{x-x'} \left(\frac{\partial}{\partial x}-\frac{\partial }{\partial x'}\right)\tilde{N}_F(x,x')  \Biggr].
\eeq 
Equating the two results, we find 
\beq
&&\frac{1}{2}\frac{d}{dx}(x\tilde{H}_g(x))+\frac{x^2}{2}\frac{d}{dx}\frac{\tilde{G}_g}{x} 
= \frac{d}{dx}\left(xG(x) -2\sum_q\int dX \Psi_{qF}\left(X+\frac{x}{2},X-\frac{x}{2}\right)\right) \nn 
&& \qquad \qquad -2\int P\frac{dx'}{x-x'} \left(\frac{\partial}{\partial x}+\frac{\partial}{\partial x'}\right)N_F(x,x') -2\int \frac{dx'}{x-x'} \left(\frac{\partial}{\partial x}-\frac{\partial}{\partial x'}\right)\tilde{N}_F(x,x') .
\eeq
Solving this for $\tilde{G}_g$ and substituting the solution into (\ref{first}), we recover  (\ref{final2}).

\bibliographystyle{apsrev}
\bibliography{ref}%

%\bibliography{ref}
%\bibliographystyle{JHEP}

\end{document}